\documentclass[11pt,a4paper]{article}
\pdfoutput=1
\usepackage{jheppub}
\usepackage{amsfonts,amsbsy,amsmath,amssymb}
\usepackage{graphicx}
\usepackage{booktabs}
\usepackage{caption}
\usepackage{subcaption}
\usepackage{multirow}
\title{The gradient flow running coupling with twisted boundary conditions.}

\author[a]{Alberto Ramos}

\affiliation[a]{NIC, DESY Platanenallee 6, 15738 Zeuthen, Germany.}
\emailAdd{alberto.ramos@desy.de}

\abstract{We study the gradient flow for Yang-Mills theories with
  twisted   boundary conditions. The perturbative 
  behavior of the energy density $\langle E(t)\rangle$ is used to
  define a running coupling at a scale given by the linear size of the
  finite volume box. We compute the non-perturbative running of the
  pure gauge $SU(2)$ coupling constant and conclude that the technique
  is well suited for further applications due to the relatively mild
  cutoff effects of the step scaling function and the high numerical
  precision that can be achieved in lattice simulations. We also
  comment on the inclusion of matter fields.
}

\preprint{%
{\flushright 
DESY 13-151\\
}}

\begin{document}
\maketitle

\section{Introduction}

Non-abelian Yang-Mills theories play a central role in our
understanding of high energy physics. Despite their apparent
simplicity (they depend on one free parameter, the coupling
constant), they have proven to produce very rich phenomena at the
quantum level. The invariance under conformal transformations present
in the classical theory is broken by quantum corrections and a typical
scale of the interactions (usually characterized by the
$\Lambda$-parameter) comes into play. The coupling constant becomes
scale dependent and asymptotic freedom~\cite{PhysRevLett.30.1343,PhysRevLett.30.1346} guarantees that at high energies
its value is small. A perturbative analysis has proven to be very
successful in making predictions in this regime, but at low energies
the coupling becomes large and non-perturbative effects set
in. Confinement characterizes the interaction in this regime 
and perturbation theory provides little information.

Finite volume renormalization schemes together with non-perturbative
numerical
simulations~\cite{Luscher:1991wu,Luscher:1992an,deDivitiis:1994yp}
play a major role in understanding how and when this 
transition from the perturbative to the non-perturbative regimes of YM
theories happens. By identifying the linear size of a finite volume box
with the renormalization scale a controlled and non-perturbative
running of the coupling constant can be carried over from the very
perturbative regime to the non-perturbative one. 

There are many practical issues that have made the Schr\"odinger
Functional~\cite{Luscher:1992an} (SF) the preferred scheme to perform
the previously mentioned program
(see~\cite{Luscher:1992zx,Luscher:1993gh,DellaMorte:2004bc,Tekin:2010mm}
for some applications). In the SF one imposes Dirichlet 
boundary conditions on the spatial components of the gauge fields at
$x_4=0,T$, while they remain periodic in the spatial directions. By
choosing appropriately the values of the fields at the time boundaries
one can define a running coupling, usually denoted $g_{\rm SF}^2$, with
many good properties from a practical point of view. It is easy to
evaluate numerically and precise, especially in the perturbative
regime.

Other schemes have been proposed: In the twisted Polyakov loop scheme
(TPL) the gauge fields are 
embedded in a torus with twisted boundary conditions and the ratio of
Poliakov loops between the twisted and periodic directions is used to
define a running coupling~\cite{deDivitiis:1994yp}. 
This scheme has some good properties, related to the manifest
invariance under translations. In particular $\mathcal O(a)$
improvement in the pure gauge case is guaranteed, and therefore one
expects smaller cutoff effects. On the other hand the observable
used to define the coupling tends to be more noisy than the SF coupling.

When one wants to study QCD-like theories, or other models with
fermions coupled to the gauge field, similar considerations apply. In
the SF scheme~\cite{Sint:1993un,Sint:1995rb} one can couple an
arbitrary number of fermions in any representation to the gauge
field. The coupling constant is defined in a similar way, and maintains
its good properties. Fermions induce additional boundary counterterms
that one needs to compute to have $\mathcal O(a^2)$ scaling. Moreover
the boundary conditions for the fermion fields typically break chiral
symmetry, and therefore one also needs
bulk improvement, although this last issue can be addressed by
modifying the boundary conditions of the fermion
fields~\cite{Sint:2010eh}. On the other hand, the TPL scheme can not
be used with an arbitrary number of fermions in the fundamental
representation. The twisted boundary conditions put a constraint on
the number of fermions that can be coupled to the gauge
field. But when this scheme can be used, it guarantees a better
scaling towards the continuum. $\mathcal O(a)$ improvement is
automatic, even with Wilson fermions, provided that one works with
massless quarks~\cite{Frezzotti:2003ni,Sint:2010eh}. Nevertheless the
observable used to define the coupling (a ratio of polyakov loops),
tends to be more noisy than the SF coupling, especially in the
non-perturbative domain. 

Recently the nice properties of the gradient
flow regarding the
renormalization of composite operators have introduced new
observables as candidates for a running coupling
definition~\cite{Luscher:2010iy,Luscher:2011bx}. By 
introducing an extra parameter (the flow 
time $t$) one defines a one parameter family of gauge fields. The
evolution of the gauge field in flow time is given by a diffusion-like
equation that drives the gauge field towards a classical solution of
the YM equations, and
therefore is a smoothing process. The key point is that the flow field
does not require any renormalization at positive flow time ($t>0$), and
correlation functions of the smooth field can be considered
as observables for a running coupling definition.

In particular the energy density of the flow field
\begin{equation}
  \langle E(t) \rangle = \frac{1}{4}\langle G_{\mu\nu}(t)G_{\mu\nu}(t)\rangle \,,
\end{equation}
where  $G_{\mu\nu}(t)$ is the field strength of the gauge field at
flow time $t$, can be used to give a non-perturbative definition of the
coupling at a scale $\mu = 1/\sqrt{8t}$. Moreover one can identify
this renormalization scale with the linear size of the box and define
a running coupling.  

This idea was first applied to the case of a periodic
box~\cite{Fodor:2012td}. The dynamics of Yang-Mills 
theories in a periodic box contains contributions from zero momentum
modes that are not Gaussian and therefore have to be treated
exactly~\cite{GonzalezArroyo:1981vw} 
(see also the review~\cite{vanBaal:1988qm} and references therein). Making
a long story short, this leads to a definition of the running coupling
that is non-analytic in $g^2_{\overline{\rm MS}}$. The author wants
to stress that there is nothing wrong with such a coupling definition,
but often one wants to make contact with perturbation theory
(i.e. when determinning the $\Lambda$ parameter). A non-analytic definition
of the coupling may make contact with perturbation theory at a larger
energy scale and looses many nice properties, like a universal 2-loop
beta function. Moreover perturbative computations in this setup are
usually more involved and difficult because one has to treat the zero
mode (toron) contribution non perturbatively. 

These difficulties can be avoided with a different choice of boundary
conditions. For example in the SF, if the boundary values are chosen
wisely, zero momentum modes become 
incompatible with the boundary conditions, and therefore these non
Gaussian modes are absent of the dynamics. A definition of a running
coupling using the gradient flow in this setup has been proposed
in~\cite{Fritzsch:2013je}. It leads to a coupling definition 
analytic in $g^2_{\overline{\rm MS}}$ and with a universal two loop
beta function.

The infamous problem of topology freezing that affects large volume
simulations~\cite{Schaefer:2010hu,Luscher:2011kk}, has recently been
shown to also affect step scaling
studies~\cite{Fritzsch:2013yxa,Luscher:2014kea}. To overcome this
problem it has been proposed~\cite{Luscher:2014kea} to use a setup
with mixed boundary conditions. Since in this scheme the fields
satisfy open boundary conditions at $x_0=T$, topological freezing is
avoided~\cite{Luscher:2012av,Luscher:2014kea}. On the other hand  
the fields satisfy Dirichlet boundary conditions at
$x_0=0$, and the absence of zero momentum modes and the
analyticity of the observable used to define the coupling is
guaranteed. 

In this paper we propose another alternative based on using twisted
boundary conditions for the gauge fields as in the TPL scheme. These
twisted boundary conditions guarantee that the action has a unique
minimum up to gauge transformations, and therefore zero momentum modes
are not present in the dynamics. Moreover the formulation is
manifestly translation invariant, since the scheme is defined in a
torus, guaranteeing the absence of $\mathcal O(a)$ effects, even when
one works with massless Wilson fermions.

This paper is organized as follows. Section~\ref{sc:tw} presents a
small review of the twisted boundary conditions. Readers interested in
more details are encouraged to look at the review~\cite{ga:torus} and
the recent paper~\cite{Perez:2013dra}. Section~\ref{sc:flow} studies the
perturbative behavior of the gradient flow with twisted boundary
conditions. Again we encourage the reader to consult the original
works~\cite{Luscher:2010iy,Luscher:2011bx} for a better understanding
of the properties of the gradient flow. Section~\ref{sc:coupling} uses
the previous results to define a non-perturbative coupling that runs
with the size of the box. Finally in section~\ref{sc:run} we compute
the running coupling for the case of an $SU(2)$ pure gauge theory, and
conclude that the modest size of cutoff effects and small variance of the
observable when using Monte Carlo techniques to compute it, make it an
interesting choice for further studies. 


\section{Twisted boundary conditions}
\label{sc:tw}

The twisted boundary conditions were first introduced by 't
Hooft~\cite{tHooft:1979uj} to characterize confinement. But for us,
they will be used simply as a tool to study
the renormalization of Yang-Mills theories. The main observation is
that the requirement for physical quantities to be periodic can be
accomplished by fields that change by a gauge
transformation under translations over a period.

\subsection{Gauge fields}

We will consider $SU(N)$
gauge fields 
in a four dimensional torus of size $L^4$. The twisted boundary
conditions are implemented by requiring the field to gauge-transform
under the displacement of a period
\begin{equation}
  A_\mu(x+L\hat \nu) = \Omega_\nu(x)A_\mu(x)\Omega_\nu^+(x) + 
  \Omega_\nu(x)\partial_\mu\Omega_\nu^+(x) \,,
\end{equation}
where $\Omega_\mu(x)$ are known as the twist matrices. The uniqueness
of $A_\mu(x+L\hat\mu+L\hat\nu)$ requires that the twist matrices have
to obey the relation 
\begin{equation}
  \label{eq:consistency}
  \Omega_\mu(x+L\hat \nu)\Omega_\nu(x) = e^{2\pi\imath n_{\mu\nu}/N}
  \Omega_\nu(x+L\hat \mu)\Omega_\mu(x)\,,
\end{equation}
where $n_{\mu\nu}$ is an anti-symmetric tensor of integers modulo $N$
called the twist tensor. It is easy to check that under a gauge
transformation, $\Lambda(x)$, the twist matrices change according to
\begin{equation}
  \Omega_\nu(x) \longrightarrow \Omega'_\nu(x) = \Lambda(x+L\hat \nu)
  \Omega_\nu(x)\Lambda^+(x)\,,
\end{equation}
but the twist tensor $n_{\mu\nu}$ remains unchanged. Therefore all the
physics of the twisted boundary conditions is contained in the twist
tensor, and the particular choice of twist matrices is 
irrelevant. One can restrict the gauge transformations to those that
leave the twist matrices unchanged. It is easy to check that the
necessary and sufficient condition for the gauge transformations is to
obey the periodicity condition
\begin{equation}
  \label{eq:gauge}
  \Lambda(x+L\hat\nu) = \Omega_\nu(x)\Lambda(x)\Omega_\nu^+(x)\,.
\end{equation}

The reader interested in knowing more about the twisted boundary
conditions is invited to consult the review~\cite{ga:torus}. Here we
will use a particular setup: we choose to twist only one plane
(the $x_1-x_2$ plane) by choosing $n_{12} = -n_{21} = 1$, while the
rest of the components of the twist tensor will be zero. This means
that our gauge connections will still be periodic in the $x_3$ and
$x_4$ directions. As we will
see, this choice guarantees that the action has a unique minimum
(modulo gauge transformations), and therefore it turns out to be a
convenient choice for perturbative studies. This is the reason why the very
same choice has been made before to define the Twisted Polyakov Loop
running coupling scheme~\cite{deDivitiis:1994yp}, or for other
perturbative studies~\cite{Luscher:1985wf}. We will closely follow the
notation and steps 
presented in~\cite{Perez:2013dra}, a reference that the reader
interested in more details should consult.

A convenient implementation of twisted boundary conditions consists in
using space-time independent twist matrices. In particular for the
periodic directions we set the twist matrices to one
\begin{subequations}
\begin{eqnarray}
  \Omega_{1,2}(x) &=& \Omega_{1,2} \\
  \Omega_{3,4}(x) &=& 1\,.
\end{eqnarray}
\end{subequations}

We will use latin indexes ($i,j,\dots=1,2$) to run over the directions in the
twisted plane, while greek indexes ($\mu,\nu,\dots=0,\dots,3$) will
run over the four space time directions. The consistency relation
Eq.~(\ref{eq:consistency}) implies the 
following condition for the twist matrices
\begin{equation}
  \Omega_1\Omega_2 = e^{2\pi\imath /N}
  \Omega_2\Omega_1.
\end{equation}
Notice that the boundary conditions for the gauge
field with this choice of the twist matrices are
\begin{equation}
  \label{eq:twalg}
  A_\mu(x+L\hat k) = \Omega_kA_\mu(x)\Omega_k^+\,,
\end{equation}
and $A_\mu=0$ is a valid connection. In fact we
will show that it is the only connection compatible with the boundary
conditions that does not depend on $x$, and therefore it is the
unique minimum of the action modulo gauge transformations. 

Eq.~(\ref{eq:twalg}) defines a generalization of the Dirac algebra. It
can be shown~\cite{ga:torus} that there is a
unique solution for the matrices $\Omega_i$ modulo similarity
transformations. Introducing the \emph{color momentum}, $\tilde p_i =
\frac{2\pi\tilde n_i}{NL}$ with $n_i=0,\dots,N-1$
it is easy to check that the $N^2$ matrices
\begin{equation}
  \label{eq:defG}
  \Gamma(\tilde p) = \frac{\imath}{\sqrt{2N}}e^{\imath \alpha(\tilde
    p)} \Omega_1^{-\tilde 
    n_2}\Omega_2^{\tilde n_1}\,,
\end{equation}
where $\alpha(\tilde p)$ are arbitrary phases, are linearly
independent and obey the relation 
\begin{equation}
  \Omega_i \Gamma(\tilde p) \Omega_i^+ = e^{\imath L\tilde p_i}
  \Gamma(\tilde p)\,.
\end{equation}
Moreover all but
$\Gamma(\tilde p=0)$ are traceless, and therefore they can be used as
a basis of the Lie algebra of the gauge group. This means that any
gauge connection can be expanded as
\begin{equation}
  A_\mu^a(x)T^a = \sum_{\tilde p}'\hat A_\mu(x,\tilde p)e^{\imath
    \tilde px}\Gamma(\tilde p).
\end{equation}
The prime over the sum means that the term $\tilde p_i=0$ is
absent in the sum, as required for a $SU(N)$ gauge group. Notice that
the coefficients $\hat A_\mu(x,\tilde p)$ are functions (not 
matrices) periodic in $x$. Therefore one can do an usual Fourier
expansion and obtain
\begin{equation}
  A_\mu^a(x)T^a = \frac{1}{L^4} \sum_{p,\tilde p}'\tilde A_\mu(p,\tilde
  p)e^{\imath 
    (p+\tilde p)x}\Gamma(\tilde p)\,,
\end{equation}
with the usual spatial momentum 
\begin{equation}
  p_\mu = \frac{2\pi n_\mu}{L}\quad (n_\mu\in \mathbb Z)\,.
\end{equation}
Finally we define the \emph{total} momentum as the sum of the color
and space momentum $P_i = p_i+\tilde p_i$, $P_{3,4} = p_{3,4}$. Noting
that any $P_\mu$ can be 
uniquely decomposed in the space momentum and color momentum degrees
of freedom we can safely write $\Gamma(P)$ instead of $\Gamma(\tilde
p)$. Our main conclusion is that any gauge connection compatible
with our choice of boundary conditions can be written as 
\begin{equation}
  \label{eq:gaugetw}
  A_\mu^a(x)T^a = \frac{1}{L^4} \sum_{P}'\tilde A_\mu(P)
  e^{\imath Px}\Gamma(P)\,.
\end{equation}
In particular the only connection that does not depend on $x$ is given
by $\tilde A_\mu(P) = 0$. In general the matrices $\Gamma(P)$ are not
anti-hermitian, but one can choose the phases 
$\alpha(P)$ of equation~(\ref{eq:defG}) so that this condition is
enforced
\begin{equation}
  \alpha(P) = \frac{\theta}{2}P_1P_2\qquad 
  \left(
    \theta = \frac{NL^2}{2\pi}
  \right)\,.
\end{equation}
In this case, the Fourier coefficients $\tilde A_\mu(P)$ satisfy the
usual relation
\begin{equation}
  \tilde A_\mu(P)^* = \tilde A_\mu(-P)\,,
\end{equation}
and the $\Gamma$ matrices are normalized according to
\begin{equation}
  {\rm Tr}\left\{ \Gamma(P)\Gamma(-P)\right\} = -\frac{1}{2}\,.
\end{equation}

We finally note that a simlar expansion is possible on the lattice,
with the only difference that the space momentum will be restricted
to the Brillouin zone. 

\subsection{Matter fields}
\label{sc:fermions}

The inclusion of matter fields interacting with gauge fields with
twisted boundary conditions is not completely straightforward. To
understand why it is better first to consider how to include 
fermion fields in the fundamental representation. Since the twist
matrices tell us how fields change under translations, one naively
expects 
\begin{equation}
  \psi(x+L\hat i) = \Omega_i\psi(x)\,,
\end{equation}
but one can easily see that this choice is not consistent, 
since the value of the field $\psi(x+L\hat i+L\hat j)$ depends on the
order in which we perform the translations due to the
non-commutativity of the twist matrices. This difficulty can be
avoided by introducing more fermions, or what usually is called a
``smell'' degree of freedom~\cite{Parisi:1984cy}. If
$\alpha,\beta=1,\dots,N_s$ are indices that run over the $N_s$ smells of
fermions, and $a,b=1,\dots,N$ run over the color degrees of freedom,
the boundary conditions of the fermions read
\begin{equation}
  \psi^a_\alpha(x+L\hat i) =
  e^{\imath\theta_i}(\Omega_i)_{ab}(\Omega^*_i)_{\alpha\beta} 
  \psi^b_\beta(x) \,.
\end{equation}
This means that a fermion smell becomes a linear combination of the
gauge transformed fermion smells under a translation. $\theta_i$
are in principle arbitrary, but introduced for 
convenience to remove the zero-momentum modes of the Dirac
operator. These phases have to be chosen
such that they are not elements of the gauge group
(i.e. $e^{\imath\theta} \not\in SU(N)$).
This choice of boundary conditions for the  
fermion fields is consistent, but they require the number of
smells to be equal to the number of colors. One can easily extend the
construction to the case when the ratio $N_s/N$ is an integer, but in
general one can not have an arbitrary number of fermions in the
fundamental representation.

On the other hand fermions in the adjoint representation transform in
the same way as the gauge fields, and therefore any number of fermions
would be compatible with the twisted boundary conditions. 

Regardless of the representation but assuming that the matter fields
are compatible with the twisted boundary conditions, $\mathcal O(a)$
improvement for massless Wilson quarks is automatically
satisfied since fields live on a torus, and the boundary conditions do
not break chiral symmetry (see~\cite{Sint:2010eh,Frezzotti:2003ni}).


\section{The gradient flow in a twisted box}
\label{sc:flow}

The gradient flow has recently proved to be an interesting tool to
study several aspects of YM
theories~\cite{Narayanan:2006rf,Luscher:2009eq,Luscher:2010iy,Luscher:2011bx,Luscher:2013cpa}. By
introducing an extra coordinate $t$, called flow time 
(not to be confused with Euclidean time $x_0$), gauge fields are
smoothed along the flow according to the equation
\begin{subequations}
  \begin{eqnarray}
    \frac{{\rm d} B_\mu(x,t)}{{\rm d}t} &=& D_\nu G_{\nu\mu} \\
    \label{eq:flow}
    B(x,0) &=& A_\mu(x)
  \end{eqnarray}
\end{subequations}
where $D_\mu = \partial_\mu + B_\mu$ is the covariant derivative and
$G_{\mu\nu} = \partial_\mu B_\nu - \partial_\nu B_\mu + [B_\mu
,B_\nu]$ is the field strength tensor. The main reason why
renormalization problems are highly simplified with the use of the
gradient flow is that correlations functions made of the flow field
$B_\mu(x,t)$ do not need renormalization at positive flow time~\cite{Luscher:2011bx}. In
particular the energy density 
\begin{equation}
  \langle E(t) \rangle = -\frac{1}{2} \langle {\rm Tr}\left\{
    G_{\mu\nu}(x,t) G_{\mu\nu}(x,t)\right\} \rangle 
\end{equation}
is a renormalized quantity at a scale $\mu = 1/\sqrt{8t}$ and can be used (at
$t>0$) to define a renormalized
coupling~\cite{Luscher:2010iy}. Moreover one can 
use a finite box to define a finite volume renormalization scheme by running
the renormalization scale with the linear size of a finite volume
box~\cite{Fodor:2012td,Fritzsch:2013je,Luscher:2014kea} 
\begin{equation}
  \mu = \frac{1}{\sqrt{8t}} = \frac{1}{cL}\,.
\end{equation}
The constant $c$ parametrizes the ratio between the
renormalization scale and the linear size of the box $L$, and
 is part of the definition of the renormalization scheme. 

Being a finite volume renormalization scheme, the boundary conditions
are relevant. The idea has been applied in a four dimensional torus
with periodic boundary conditions~\cite{Fodor:2012td}, with
Schr\"odinger functional boundary
conditions~\cite{Fritzsch:2013je}, and also with mixed boundary
conditions~\cite{Luscher:2014kea}. In this work we will give a 
definition of a running coupling in a four dimensional torus with
twisted boundary conditions, but first we need to study the
perturbative behavior of $\langle E(t) \rangle$ in a twisted box.

\subsection{Perturbative behavior of the gradient flow in a twisted box: continuum}

\subsubsection{Generalities and gauge fixing}

We are interested in the perturbative expression for  $\langle E(t)
\rangle$, and in order to avoid some difficulties in the definition of
propagators, it turns out to be convenient to fix the gauge of the
flow field $B_\mu(x,t)$. This can be achieved by studying the modified
flow equation
\begin{equation}
    \frac{{\rm d} B_\mu^{(\alpha)}(x,t)}{{\rm d}t} = D_\nu^{(\alpha)}
    G_{\nu\mu}^{(\alpha)}(x,t) +  
  \alpha D_\mu^{(\alpha)}\partial_\nu B_\nu^{(\alpha)}(x,t) \,.
\end{equation}
The superscript ${(\alpha)}$ recalls that covariant derivatives and field
strength are made of the modified flow field $B_\mu^{(\alpha)}(x,t)$,
solution of the previous equation. A solution of this modified flow
equation $B_\mu^{(\alpha)}(x,t)$ can be transformed in a solution of the
original flow equation~(\ref{eq:flow}) by a time dependent gauge
transformation~\cite{Luscher:2011bx}
\begin{equation}
  B_\mu = \Lambda B_\mu^{(\alpha)}\Lambda^{-1} + 
  \Lambda \partial_\mu
  \Lambda^{-1} \,,
\end{equation}
where 
\begin{equation}
  \frac{{\rm d} \Lambda}{{\rm d}t} =
  \alpha \Lambda \partial_\mu B_\mu \,;\quad
  \Lambda\big|_{t=0} = 1\,.
\end{equation}

Therefore gauge invariant quantities are independent of $\alpha$. Note
that the previously defined gauge transformation 
$\Lambda(x)$ belongs to the restricted set of gauge transformations
that leave the twist matrices invariant (see
equation~(\ref{eq:gauge})), and the boundary conditions of
$B_\mu^{(\alpha)}$ are also independent of $\alpha$. 

\subsubsection{Flow field and energy density to leading order}

The particular choice $\alpha=1$ simplify the computations, and we
will use it for the rest of this section. The modified flow equation
reads in this case
\begin{equation}
  \label{eq:flowmd}
  \frac{{\rm d} B_\mu}{{\rm d}t} = D_\nu G_{\nu\mu} +
  D_\mu\partial_\nu B_\nu \,.
\end{equation}
In perturbation theory one re-scales the gauge potential with the bare
coupling $A_\mu \rightarrow g_0A_\mu$, and the flow field has an
asymptotic expansion in the bare coupling
\begin{equation}
  \label{eq:flowfg0}
  B_\mu(x, t) = \sum_{n=1}^{\infty} B_{\mu,n}(x, t) g_0^n \,.
\end{equation}
To leading order our flow equation~(\ref{eq:flowmd}) is just the heat
equation
\begin{eqnarray}
  \label{eq:flowlo}
   \frac{{\rm d} B_{\mu,1}(x,t)}{{\rm d}t} &=& \partial_\nu^2
   B_{\mu,1}(x,t) \\
   B_{\mu,1}(x,0) &=& A_{\mu}(x)\, ,
\end{eqnarray}
expanding $B_{\mu,1}(x,t)$ in our preferred basis~(\ref{eq:gaugetw}) one
can easily solve~(\ref{eq:flowlo}) and obtain
\begin{equation}
  B_{\mu,1}(x,t) = \frac{1}{L^4}
  \sum_P' e^{-P^2t} \tilde A_\mu(P) e^{\imath Px}
  \Gamma(P)\,.
\end{equation}

Finally our observable of interest also has an expansion in powers of
$g_0$
\begin{equation}
  \langle E(t)\rangle = -\frac{1}{2}\langle
  {\rm Tr}\{G_{\mu\nu}(x, t)G_{\mu\nu}(x,t)\}\rangle = \mathcal E(t) + \mathcal
  O(g_0^4)\,.
\end{equation}
One can easily obtain 
\begin{eqnarray}
  \mathcal E(t) &=& \frac{g_0^2}{2}\langle 
  \partial_\mu B_{\nu,1}\partial_\mu B_{\nu,1} - \partial_\mu
  B_{\nu,1}\partial_\nu B_{\mu,1} 
  \rangle  \\
  \nonumber
  &=& \frac{-g_0}{2L^8}\sum_{P,Q}'e^{-(P^2+Q^2)t}e^{\imath (P+Q)x} 
  \left( P_\alpha Q_\alpha\delta_{\mu\nu} -
    P_\mu Q_\nu\right) \langle \tilde A_\mu(P)\tilde A_\nu(Q)\rangle 
  {\rm Tr}(\hat\Gamma(P)\hat\Gamma(Q))\,.
\end{eqnarray}
Finally using the expression for the gluon propagator
\begin{equation}
  \langle \tilde A_\mu(P)\tilde A_\nu(Q) \rangle =
  L^4 \delta_{P_\alpha, -Q_\alpha} \frac{1}{P^2}
  \left[ \delta_{\mu\nu} - (1-\lambda^{-1})\frac{P_\mu P_\nu}{P^2}\right]
  \frac{1}{{\rm Tr}({\Gamma(-P)\Gamma(P)})} + \mathcal O(g_0^2)
\end{equation}
one gets
\begin{equation}
  \label{eq:et}
  \mathcal E(t) = 
  \frac{3g_0^2}{2L^4}\sum_P' e^{-2P^2t}\,.
\end{equation}

\subsection{Perturbative behavior of the gradient flow in a twisted box: lattice}

When defining the gradient flow in the lattice one has to make several
choices. These basically correspond to the particular discretizations
of the action whose gradient is used to define the flow, as well as
the discretization of the energy density and the choice of action that
one simulates (i.e. Wilson/improved actions). 

First we will analyze the popular case where the Wilson action is
simulated, and one uses the same action to define the flow (in this
case is called the Wilson flow). The clover definition of the
observable has been a typical choice~\cite{Luscher:2010iy} for a 
discretization of
the energy density. Later we will comment on the general case. 

\subsubsection{Generalities and gauge fixing}

On the lattice the gradient flow is substituted by a discretized
version. There are several possibilities: one can use the Wilson
action  
\begin{equation}
  S_w(V) = \frac{1}{g_0^2} \sum_{\rm p} {\rm Re}\{{\rm Tr}(1-U_{\rm p})\}
\end{equation}
where the sum runs over the oriented plaquettes, and define the flow
equation by equating the time derivative of the links with
the gradient of the Wilson action
\begin{equation}
  \label{eq:flowlat}
  a^2\partial_t V_\mu(x,t) = -g_0^2 \{T^a\partial_{x,\mu}^a S_w(V)\}
  V_\mu(x,t) \,,  \qquad V_\mu(x,0) = U_\mu(x)  \,.
\end{equation}
In this case the gradient flow is usually referred as the Wilson
flow. Some explanations of our notation are in order. 
If
$f(U_\mu(x))$ is an arbitrary function of the link variable
$U_\mu(x)$, the components of its Lie-algebra valued derivative
$\partial_{x,\mu}^a $  
are defined as 
\begin{equation}
   \partial_{x,\mu}^a f(U_\mu(x)) = \left.\frac{ {\rm d} f(e^{\epsilon
            T^a}U_\mu(x))}{ {\rm d}\epsilon} \right|_{\epsilon=0}\,. 
\end{equation}
In perturbation theory one is interested in a neighborhood of the
classical vacuum configuration. In this neighborhood the lattice  
fields $U_\mu(x)$ and $V_\mu(x,t)$ are parametrized as follows:
\begin{align}
  U_\mu(x)   &= \exp\{ag_0 A_\mu(x)\}   \;, &
  V_\mu(x,t) &= \exp\{ag_0 B_\mu(x,t)\} \;.
\end{align}

Again it is convenient to study a modified flow equation where the
gauge degrees of freedom are damped. We will consider
\begin{equation}
  \label{eq:flowlatmd}
  a^2\partial_t V_\mu^\Lambda(x,t) = g_0^2 \left\{ 
        -\big[ T^a\partial_{x,\mu}^a S_w(V^\Lambda) \big] 
        + a^2\hat D_\mu^{\Lambda}\big[\Lambda^{-1}(x,t)\dot \Lambda(x,t)\big] 
                                           \right\} V_\mu^\Lambda(x,t) \,,
\end{equation}
with $V_\mu^\Lambda(x,0) = U_\mu(x)$ and the forward lattice covariant
derivative 
$\hat D_\mu^{\Lambda}$ acting on Lie-algebra valued functions according to
\begin{equation}
  \hat D_\mu f(x) = \frac{1}{a}\left[
    V_\mu(x,t)f(x+\hat\mu)V_\mu^{-1}(x,t) - f(x)
  \right] \,.
\end{equation}

Solutions of the modified and original flow equations are related by a
gauge transformation
\begin{equation}
  V_\mu(x,t) = \Lambda(x,t)V_\mu^\Lambda(x,t)\Lambda^{-1} (x+\hat\mu,t)\,.
\end{equation}
The most natural choice for $ \Lambda(x,t)$ is the same functional
used for gauge fixing 
\begin{equation}
  \label{eq:lam}
  \Lambda^{-1}\frac{{\rm d} \Lambda}{{\rm d}t} = \alpha
  \hat\partial^*_\mu B_\mu(x,t)\,,\qquad 
  \Lambda\big|_{t=0} = 1\,.
\end{equation}
where $\hat \partial, \hat \partial^*$ denote the forward/backward 
finite differences. We again note that the boundary conditions of
$V_\mu^\Lambda(x,t)$ are independent of $\alpha$, since $\Lambda(x,t)$
belongs to the restricted class of gauge transformations that leave
the twist matrices unchanged.

\subsubsection{Flow field and energy density to leading order}

Again the choice $\alpha=1$ turns out to be convenient and we
will stick to it from now on.

The modified flow equation reads
\begin{equation}
  a^2\partial_t V_\mu(x,t) = g_0^2 \left\{ -[T^a\partial_{x,\mu}^a
      S_w(V)] + a^2\hat D_\mu(\hat\partial_\nu^* B_\nu ) 
      \right\}
  V_\mu(x,t) \,,  \qquad V_\mu(x,0) = U_\mu(x) \,.
\end{equation}
The flow field can be expanded in powers of $g_0$
(equation~(\ref{eq:flowfg0})) and to first order in $g_0$ we have
\begin{equation}
  \label{eq:wflowlato1}
  \partial_t B_{\mu,1}(x,t) = \hat \partial_\nu\hat\partial_\nu^*
  B_{\mu,1}(x,t) \,.
\end{equation}
Expanding the flow field in our favorite Lie-algebra basis
(equation~(\ref{eq:gaugetw})) one can write the solution to the
previous equation
\begin{equation}
  B_{\mu,1}(x,t) = \frac{1}{L^4}\sum_P' e^{-\hat P^2t} \tilde A_\mu(P)
  e^{\imath Px} \Gamma(P)\,,
\end{equation}
where 
\begin{equation}
  \hat P_\mu = \frac{2}{a}\sin\left(a\frac{P_\mu}{2}\right)
\end{equation}
is the usual lattice momentum.

We can choose among different discretizations for the energy
density. The most popular one consists in using the clover definition
for $G_{\mu\nu}(x,t)$~\cite{Luscher:2010iy}. To leading order we have
\begin{eqnarray}
  \nonumber
  \hat G_{\mu\nu}(x,t) &=& \frac{g_0}{2}\,\mathring\partial_\mu\left[B_{\nu,1}(x,t) + 
    B_{\nu,1}(x-\hat \nu,t)\right] \\
  &-&
  \frac{g_0}{2}\,\mathring\partial_\nu\left[B_{\mu,1}(x,t) + 
    B_{\mu,1}(x-\hat \mu,t)\right] + \mathcal O(g_0^2) \,,
\end{eqnarray}
where $\mathring\partial_{\mu} = \tfrac{1}{2}(\hat \partial_\mu +
\hat \partial^*_\mu)$ is the symmetric finite difference. The energy
density computed with the clover definition for the field strength
tensor reads 
\begin{equation}
  \langle E^{\rm cl}(t)\rangle = -\frac{1}{2}\langle {\rm Tr}\{ \hat
  G_{\mu\nu} \hat G_{\mu\nu}\} \rangle = \mathcal E^{\rm cl}(t, a/L) +
  \mathcal O(g_0^2) 
\end{equation}
Using the definitions
\begin{subequations}
\begin{eqnarray}
  \mathring P_\mu &=& \frac{1}{a}\sin\left(aP_\mu\right)\,,\\
  C_\mu &=& \cos\left(a\frac{P_\mu}{2}\right)\,,
\end{eqnarray}
\end{subequations}
and the lattice gluon propagator, one can easily obtain
\begin{equation}
  \label{eq:etcl}
  \hat{\mathcal E}^{\rm cl}(t, a/L) = 
  \frac{g_0^2}{2L^4}\sum_{P}' e^{-2\hat P^2t}
  \frac{\mathring P^2 C^2 - (\mathring P_\mu C_\mu)^2}{\hat P^2}\,.
\end{equation}

\subsubsection[Some comments on different
  discretizations]{Some comments on different
  discretizations\protect\footnote{The author wants to thank S. Sint for his
    help in understanding the points discussed in this section.}}
\label{sc:disc}

In general the lattice computation of the leading order behavior of
the energy density involves several choices of discretization: the
action that one simulates (labelled (a)), the action whose gradient
defines the flow evolution (labelled (f)), and finally the
discretization used to compute the observable (labelled (O)). To
leading order, these three choices can be expressed as 
choice of ``actions'' 
\begin{subequations}
  \begin{eqnarray}
    S_a[\tilde A_{\mu}] &=& \frac{1}{4L^4}\sum_{P}' \tilde A_\mu(-P)
    K_{\mu\nu}^{(a)}(P) \tilde 
    A_\nu(P) + \mathcal O(g_0^2)\,,\\
    S_f[\tilde A_{\mu}] &=& \frac{1}{4L^4}\sum_{P}' \tilde A_\mu(-P)
    K_{\mu\nu}^{(f)}(P) \tilde 
    A_\nu(P) + \mathcal O(g_0^2)\,, \\
    S_O[\tilde A_{\mu}] &=& \frac{1}{4L^4}\sum_{P}' \tilde A_\mu(-P)
    K_{\mu\nu}^{(O)}(P) \tilde 
    A_\nu(P) + \mathcal O(g_0^2) \,.
  \end{eqnarray}
\end{subequations}

The matrices $K^{(a)}$ and $K^{(f)}$ may (and should) contain a gauge
fixing part, but not the one corresponding to the observable
$K^{(O)}$. In this way final results will be independent of the
choices of gauge.
The inverse of the $K_{\mu\nu}^{(a)}$ defines the lattice gluon propagator
\begin{eqnarray}
  \langle A_\mu(-P)A_\nu(P)\rangle &=& D_{\mu\nu}(P)\,, \\
  K_{\mu\alpha}^{(a)}(P)D_{\alpha\nu}(P) &=& \delta_{\mu\nu}\,.
\end{eqnarray}

Using this notation it is trivial to obtain the form of the flow field
to leading order
\begin{equation}
  \tilde B_{\mu,1}(P) = \left( \exp\{-t K^{(f)}(P)\}\right)_{\mu\nu}
  \tilde A_\nu(P) = H_{\mu\nu}(t,P) \tilde A_\nu(P) \,, 
\end{equation}
and noting that the reality of the action requires that $H^+(t,P) =
H(t,-P)$,  we can write the expression of the energy density to
leading order as 
\begin{eqnarray}
  \mathcal E(t,a/L) &=& g_0^2 \langle S_O[\tilde
  B_{\mu,1}]\rangle \\ 
  &=& \frac{g_0^2}{2L^4} \sum_P' {\rm Tr}\{ H^+(t,P)K^{(O)}(P)H(t,P) 
  D(P)\}\,.
\end{eqnarray}

This formula allows an easy evaluation of the energy density, to
leading order in perturbation theory, for any choice of
discretizations. One general point that one can make is that if one
uses the Wilson flow the matrix $H(t,P)$ can be chosen to be
proportional to the identity (by an appropriate gauge choice), and
therefore commutes with any other matrix. Moreover if the 
action that one simulates is the same as the one that we use to
compute the observable, the product of matrices $DK^{O}$ together with
the trace simply result in a factor 3, and therefore one obtains
\begin{equation}
  \mathcal E(t,a/L) = \frac{3g_0^2}{2L^4} \sum_P' e^{-2t\hat P^2}\,.
\end{equation}

This means that without changing the flow, improving
the action and the observable leads to exactly the same cutoff effects
than if one does not improve anything (to leading order). 

\subsection{Tests}

In order to check the previous computations one can perform several
consistency checks. First it is obvious that the continuum result
(equation~(\ref{eq:et})) is recovered from the lattice one 
(equation~(\ref{eq:etcl})) if one takes the limit $a/L \rightarrow
0$. In the infinite volume limit boundary conditions are irrelevant,
and therefore for $L\rightarrow \infty$ one should recover the result
of~\cite{Luscher:2010iy} that reads
\begin{equation}
  \mathcal E^{(L=\infty)}(t) = \frac{3g_0^2(N^2-1)}{128\pi^2 t^2}\,.
\end{equation}

This result is reproduced from our expression
equation~(\ref{eq:et}) by simply noting that 
\begin{equation}
  P_\mu = \frac{2\pi}{L}\left( n_\mu + \frac{\tilde n_\mu}{N}\right)\,,
\end{equation}
and therefore 
\begin{equation}
  \frac{1}{L^4} \sum_P' \xrightarrow[L\rightarrow \infty]{}
  \frac{1}{(2\pi)^4}\sum_{\tilde p_i}' \int_{-\infty}^\infty {\rm d}^4P\,.
\end{equation}
Finally recalling that there are $N^2-1$ terms in the sum over $\tilde
p_i$ (the term $\tilde p_i=0$ is explicitly excluded) one obtains
\begin{equation}
  \mathcal E(t)\xrightarrow[L\rightarrow \infty]{}
  \frac{3g_0^2}{32\pi^4}\sum_{\tilde p_i}' \int_{-\infty}^\infty {\rm d}^4P
  e^{-2P^2 t} = \frac{3g_0^2(N^2-1)}{128\pi^2 t^2}\,.
\end{equation}

\begin{figure}
  \centering
  \begin{subfigure}[b]{0.45\textwidth}
    \includegraphics[width=\textwidth]{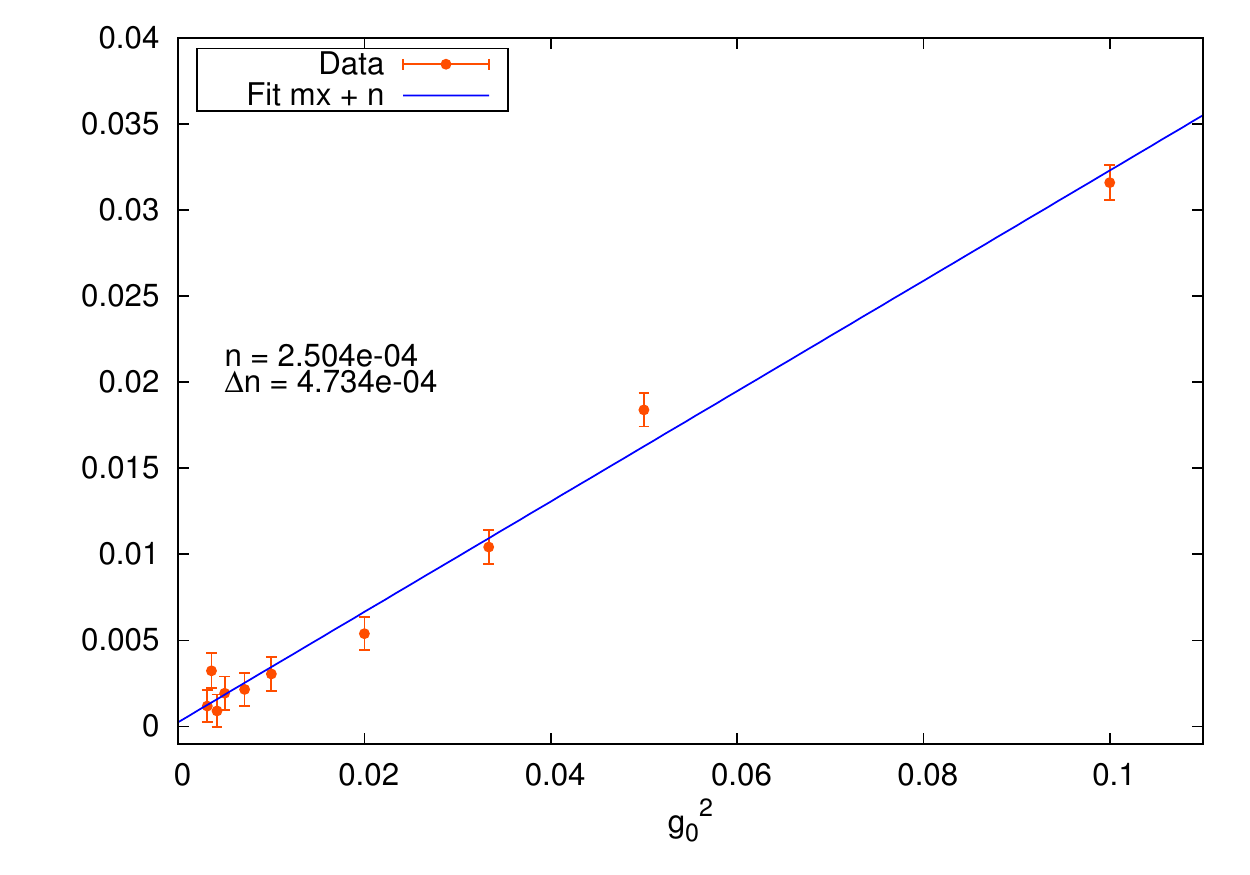}
    \caption{Fit for $L=4^4$ and $c=0.3$.}
  \end{subfigure}
  \begin{subfigure}[b]{0.45\textwidth}
    \includegraphics[width=\textwidth]{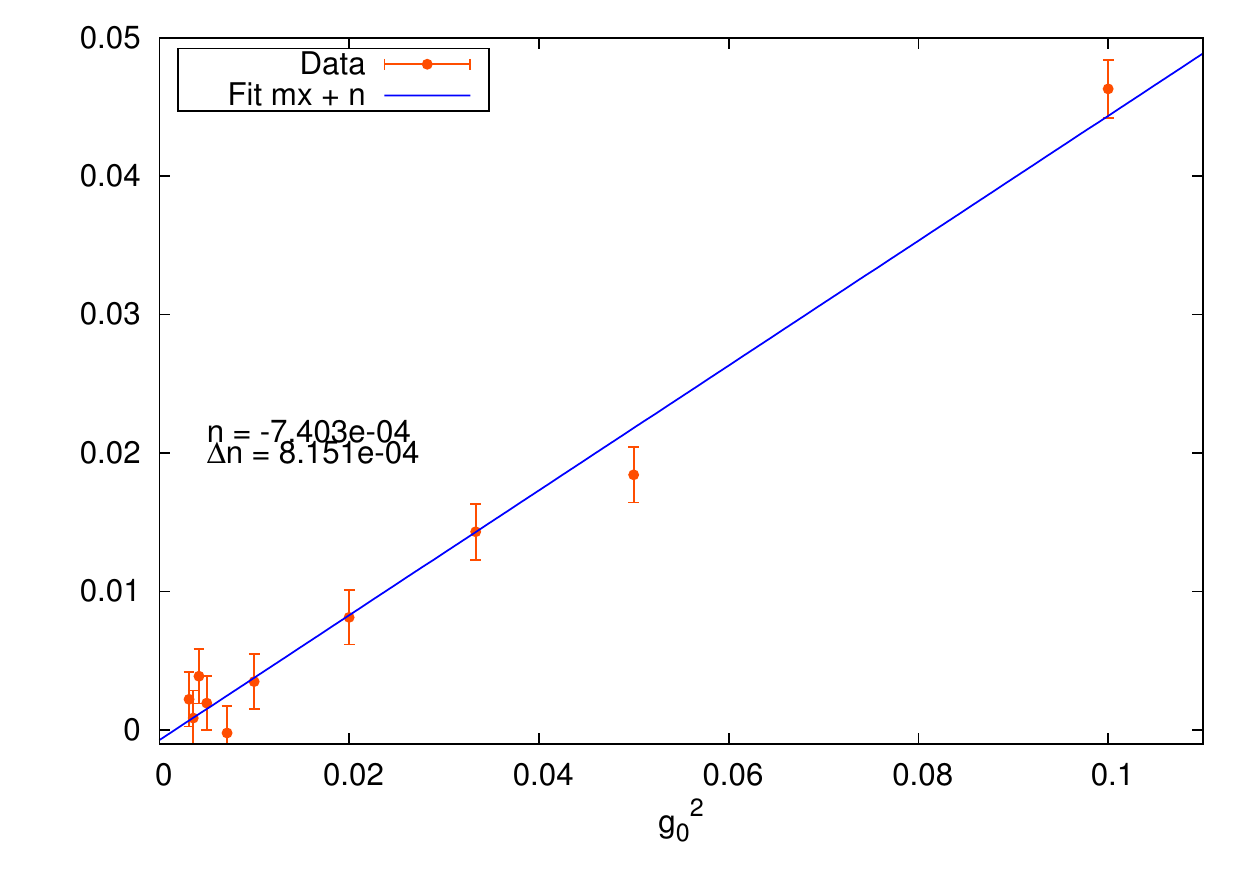}
    \caption{Fit for $L=6^4$ and $c=0.5$}
  \end{subfigure}

  \caption{Some representative fits to the large-$\beta$
    simulations. The plots show the function $R(g_0, t)$ at fixed
    $t=c^2L^2/8$ versus $g_0^2$.} 
  \label{fig:fit}
\end{figure}

To check the lattice computations we have performed some dedicated
pure gauge lattice 
simulations. We use the plaquette action of an $SU(2)$ gauge theory in
two different volumes $L/a=4^4$ and $L/a= 
6^4$. We collect $10,000$ measurements of $\langle E^{\rm
  cl}(t)\rangle$ for different values of $t$ and $\beta = 2/g_0^2 =
40, 80,  120,  200,  400,  560,  800,  960,  1120, 1280$. In these
large-$\beta$ simulations the 
measured $\langle E^{\rm cl}(t)\rangle$ should
reproduce the perturbative expression
(equation~(\ref{eq:etcl})). Being more precise, we will study
numerically the
quantity
\begin{equation}
  R(g_0, t) = \frac{\langle E^{\rm cl}(t)\rangle - \mathcal E^{\rm
      cl}(t)}{\mathcal E^{\rm cl}(t)}\,.
\end{equation}
We expect that $R(g_0,t) = \mathcal O(g_0^2)$, and therefore by fitting
the data from the simulations to a linear behavior 
\begin{equation}
  R(g_0, t)  = m(t)g_0^2 + n(t)
\end{equation}
one should obtain an intercept $n(t)$ compatible with zero within
errors. Indeed this is the case, for different values of $t$ and $L$,
as the reader can check in table~(\ref{tab:fit}). A couple of
representative fits are shown in the figure~\ref{fig:fit}.

\begin{table}
  \centering
  \begin{tabular}{ll|lllll}
    \toprule
    $V$ &  & \multicolumn{5}{c}{$c$} \\
    & & 0.30 &0.35 &0.40 &0.45 &0.50 \\
    \midrule
    \multirow{3}{*}{$4^4$}
    & $\chi^2/{\rm ndof}$ & 1.4 & 1.5 & 1.6 & 1.7 & 1.7 \\
    & $\mathcal E(t,a/L)$ & $1.37\times 10^{-1}$ & $9.76\times
    10^{-2}$ & $6.58\times 10^{-2}$ & $3.22\times 10^{-2}$& 
    $2.53\times 10^{-2}$ \\
    & $n\times 10^4$ &$2.5(4.7)$ & $3.4(6.2)$ & $4.8(8.1)$ & $6.8(10)$
    & $9.5(12)$ \\
    \midrule
    \multirow{3}{*}{$6^4$}
    & $\chi^2/{\rm ndof}$ & 0.34 & 0.44 & 0.62 & 0.83 & 1.03 \\
    & $\mathcal E(t,a/L)$ & $3.78\times 10^{-2}$ & $2.20\times
    10^{-2}$ & $1.39\times 10^{-2}$ & $7.82\times 10^{-3}$& 
    $5.45\times 10^{-3}$ \\
    & $n\times 10^4$ & $-1.8(1.6)$ & $-3.1(2.5)$ & $-4.5(3.9)$ &
    $-6.0(5.9)$ & $-7.4(8.2)$ \\
    \bottomrule
  \end{tabular}
  \caption{Results of a linear fit of $R(g_0,t)$ for different lattice
    sizes and different flow times $t$. In the table the flow time is
    parametrized by $c=\frac{\sqrt{8t}}{L}$. As one can observe all
    fits have a good fit quality and the intercept of the fit (given
    by $n$) is compatible with zero within errors.}
  \label{tab:fit}
\end{table}


\section{Running coupling definition}
\label{sc:coupling}

The computations of the previous section guarantee that 
\begin{equation}
  t^2\langle E(t)\rangle = g_0^2 
  \left[\frac{3t^2}{2L^4} \sum_P' e^{-2P^2t} \right]
  + \mathcal O(g_0^4)\,.
\end{equation}
On the other hand the properties of the gradient flow ensures 
that $t^2\langle E(t)\rangle$ is a renormalized observable defined at
a scale $\mu = 1/\sqrt{8t}$. This suggests that one can use $t^2\langle
E(t)\rangle$ for a non-perturbative coupling definition. Moreover if
one keeps the product of the renormalization scale and the linear size
of the box fixed (i.e. $\mu L = 1/c = {\rm constant}$) the coupling
will depend on no scale other than the linear size of the box, and
therefore will be ideal for finite size scaling. 

In full glory our coupling definition reads
\begin{equation}
  g_{\rm TGF}^2(L) = \mathcal N^{-1}_T(c)t^2\langle E(t) \rangle \Big|_{t=c^2L^2/8}
\end{equation}
where 
\begin{equation}
    \mathcal N_T(c) =
  \frac{3c^4}{128}\sum_P' e^{-\frac{c^2L^2}{4}P^2}
  = \frac{3c^4}{128}\sum_{n_\mu=-\infty}^{\infty}
  {\sum_{\tilde n_i=0}^{N-1}}' 
  e^{-{\pi^2c^2}(n^2 + \tilde n^2/N^2 + 2\tilde n_i n_i/N)}\,.
\end{equation}

This coupling has a perturbative expansion
\begin{equation}
  g_{\rm TGF}^2(L) = g_{\rm \overline{MS}}^2 + \mathcal O(g_{\rm \overline{MS}}^4)
\end{equation}
and a universal two loop $\beta$-function
\begin{equation}
  L\frac{\partial g_{\rm TGF}^2}{\partial L} = -\frac{\beta_0}{16\pi^2}
  g_{\rm TGF}^4 - \frac{\beta_1}{\left(16\pi^2\right)^2}
  g_{\rm TGF}^6 + \mathcal O(g_{\rm TGF}^8)
\end{equation}
where the universal coefficients, for the case of an $SU(N)$ YM
theory are given by
\begin{equation}
    \beta_0 = \frac{11N}{3}\,,\qquad
    \beta_1 = \frac{34N^2}{3}\,.
\end{equation}

We point out that the same coupling definition is valid if one includes
any number of fermions in any representation, as soon as they are
allowed by the twisted boundary conditions (more details in
section~\ref{sc:fermions}).

\subsection{Cutoff effects in the twisted running coupling}

\begin{figure}[h]
  \centering
  \includegraphics[width=\textwidth]{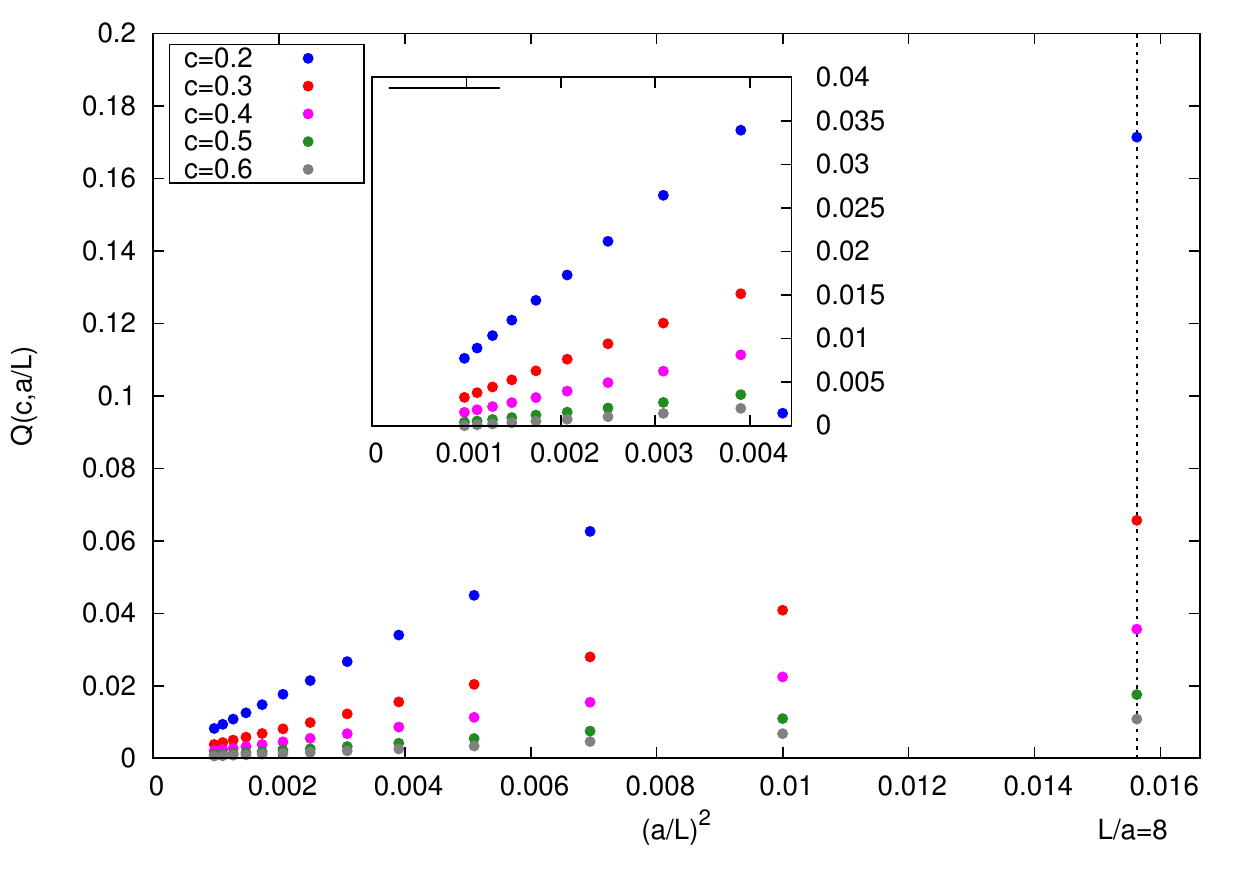}
  \caption{Cutoff effects to leading order of perturbation theory in
    the twisted gradient flow coupling. As we 
  see, for $c\in[0.3-0.5]$ cutoff effects are below the 7\% for an
  $L/a=8$ lattice.}
  \label{fig:cut1}
\end{figure}

The comparison of the lattice and the continuum computations of
$\mathcal E(t)$ can give us an idea of the size of cutoff effects (to
leading order in $g_0^2$) of the twisted gradient flow coupling. We
are going to study in detail the case of lattice simulations using the
Wilson action, the Wilson flow, and the clover definition for the
observable. If we
define 
\begin{equation}
  \hat{\mathcal N}_T(c,a/L) = 
  \frac{c^4}{128}\sum_P' e^{-\frac{c^2L^2}{4}\hat P^2}
  \frac{\mathring P^2 C^2 - (\mathring P_\mu C_\mu)^2}{\hat P^2}\,,
\end{equation}
the quantity
\begin{equation}
  Q(c, a/L) = \left|\frac{\hat{\mathcal N}_T(c,a/L) - \mathcal N_T(c)}
{{\mathcal N}_T(c)}\right|\,,
\end{equation}
quantifies to leading order the size of cutoff effects as a function of the
lattice size and the scheme parameter $c$. A global picture of cutoff
effects for the groups $SU(2)$ and $SU(3)$ 
can be seen in the figure~\ref{fig:cut1}. 

These figures may lead to the conclusion that a large value of $c$ is
optimal. But from the point of view of lattice simulations, it is
known~\cite{Fritzsch:2013je} that larger values of $c$ lead to larger
statistical errors when computing the coupling via lattice
simulations. For the typical lattice sizes ($L/a\sim 10-20$) that one
uses in step scaling studies the values $c\in[0.3,0.5]$ seem reasonable. 

\subsection{Improved coupling definition}

If one is computing $t^2\langle E(t)\rangle$ non-perturbatively via
lattice simulations, and one is using the Wilson action, the Wilson
flow and the clover observable for the evaluation of the energy
density observable, one can alternatively define the coupling via
\begin{equation}
\label{eq:latcou}
  g_T^2(L) = \hat{\mathcal N}_T^{-1}(c,a/L)t^2\langle E(t) \rangle
  \Big|_{t=c^2L^2/8} \,.
\end{equation}
This last coupling definition has the same properties, but one expects
an improved scaling towards the continuum limit, since the leading
order cutoff effects $\propto g_0^2$ have been removed thanks to
the lattice factor $\hat{\mathcal N}_T(c,a/L)$.

In a similar way, any choice of discretizations that define a coupling
can be normalized with a factor computed on the lattice
(cf. section~\ref{sc:disc}), leading to an improved scaling towards
the continuum.


\section{$SU(2)$ running coupling}
\label{sc:run}

In this section we will compute the running coupling
in $SU(2)$ pure gauge theory to test if the twisted coupling
definition is applicable for step scaling studies.

We will first recall the general strategy, introduced
in~\cite{Luscher:1991wu},  of the recursive procedure
involved in the computation of the running coupling. The process
starts with a non-perturbative coupling definition that depends on no
scale other than the linear size of a finite-volume box. Of course
in our case this role will be played by the twisted gradient flow
coupling $g^2_{\rm TGF}(L)$. The step scaling function tells us how
much the coupling changes when the renormalization scale is changed by
a factor $s$
\begin{equation}
  \sigma_s(g^2(L)) = g^2(sL)\,.
\end{equation}
Therefore it is a discrete version of the $\beta-$function. The value
$s=1/2$ is a typical choice, and is the one we will use for now on,
although the basic idea is the same for any other value. If one
knows how much is the coupling at a renormalization scale that
corresponds to a large volume (lets call it $L_{\rm max}$), and one
knows the step scaling function, then
one can obtain the value of the coupling at scales $L_{\rm
  max}/2^k$ for $k=1,2,\dots$. 
Eventually, one will reach a very small box size (very large
energy scale), where asymptotic freedom guarantees
that one can safely make contact with perturbation theory.

To compute the step scaling function numerically one starts by
measuring the coupling on some lattice of size $L/a$. 
The step scaling function is easily obtained by
simply measuring the coupling on a lattice half as big ($L/(2a)$)
while keeping the rest of the bare parameters constant. The step
scaling function computed in this naive way will carry an implicit
dependence of the lattice spacing (the cutoff), and therefore defines
a lattice step scaling function 
\begin{equation}
  \Sigma(g^2(L), a/L)\,.
\end{equation}
In order to obtain the continuum step scaling function
$\sigma(g^2(L))$, one simulates several pairs of lattices and takes
a continuum limit:
\begin{equation}
  \sigma(u) = \lim_{a/L\rightarrow 0} \Sigma(u,a/L)\,.
\end{equation}

\subsection{Numerical computation of the step scaling function and
  running coupling}

\subsubsection{Simulation details}

We will simulate $SU(2)$ YM theory using the Wilson action
\begin{equation}
  S = \frac{\beta}{4}\sum_{\rm p} {\rm Tr}\left\{ 1-U_{\rm p}\right\}
\end{equation}
where the sum runs over all oriented plaquettes. We simulate lattices
of size $L/a=20, 24, 30, 36$, and in order 
to compute the step scaling function also lattices of half this
size ($L/a=10, 12, 15, 18$). The range of values of $\beta$ (between 2.75
and 12.0) translate to renormalized couplings
$g_{\rm TGF}^2(L)$ between 7.5 and 0.6 (for $c=0.3$), enough to cover both the
non-perturbative and perturbative regions of the 
theory. Appendix~\ref{ap:values} collects the values of the $g^2_{\rm
  TGF}(L)$ of our simulations. 

We will use a combination of heatbath~\cite{Creutz:1980zw,Fabricius:1984wp,Kennedy:1985nu} and
overrelaxation~\cite{Creutz:1987xi} as suggested
in~\cite{Wolff:1992nq}. In particular we 
choose to do one heatbath sweep followed by $L/a$ overrelaxation
sweeps. Since measuring the coupling (i.e. integrating the flow
equations) is numerically more expensive 
than the Monte Carlo updates, we repeat this process 50 times between
measurements. 

In total we collect 2048 measurements of the coupling for each lattice
size, each value of $\beta$, and several values of
$c\in[0.3,0.5]$. These measurements are collected in $N_r$ 
parallel runs (replica) of length $N_{\rm MC}$ each so that $N_r\times
N_{\rm MC} = 2048$. 
We check that there are no autocorrelation between
measurements (i.e. $\tau_{\rm int}=0.5$ within errors), even for our
larger lattices and larger values of $c$. We conclude that we can
safely consider the measurements independent. 

The Wilson flow equations are integrated using the adaptive step size
integrator described in appendix D of~\cite{Fritzsch:2013je}. With
this scheme we 
make sure that the integration error in each step is not larger than
$10^{-6}$. 

\subsubsection{Data analysis}

For each $L/a$ we have computed the value of the twisted gradient flow
coupling at different values of
$\beta$ (we call it $g^2_{\rm TGF}(\beta;L/a)$). These data are fitted to a
Pad\'e-like ansatz
\begin{equation}
  \label{eq:pade}
  g^2_{\rm TGF}(\beta;L/a) = \frac{4}{\beta} \quad
  \frac{\sum_{n=0}^{M-1} a_n\beta^n + \beta^M}
  {\sum_{n=0}^{M-1} b_n\beta^n + \beta^M}\,.
\end{equation}
This fit imposes the one-loop constraint to the data (i.e. $g^2_{\rm
  TGF}(\beta;L/a) \rightarrow 4/\beta$ at large $\beta$), and has
a total of $2M$ free fit parameters. 

Alternatively, and to estimate the dependence of our results on the 
choice of functional form used to fit the data, we use a different Pad\'e
inspired functional form
\begin{equation}
  \label{eq:taylor}
  g^2_{\rm TGF}(\beta;L/a) = \frac{4}{\beta} \quad
  \frac{1}
  {1 + \sum_{n=1}^{M} c_n/\beta^n}\,,
\end{equation}
that also ensures the correct one-loop behavior at large $\beta$.

We obtain good fits ($\chi^2/{\rm
ndof}\sim 0.6-1.9$) with $M=2$ when using the functional form of
Eq.~\ref{eq:pade} to fit the lattice data (i.e. 4 fitting
parameters). When using the functional form of Eq.\ref{eq:taylor} we
need $M=4$ to accurately describe the data on the small lattices
($L/a=10,12$) and $M=6$ for the larger ones
($L/a=15,18,20,24,30,36$). It is important to 
stress that the data are statistically uncorrelated, since they
correspond to different simulations. 

\begin{figure}
  \centering
        \begin{subfigure}[t]{0.45\textwidth}
          \centering
          \includegraphics[width=\textwidth]{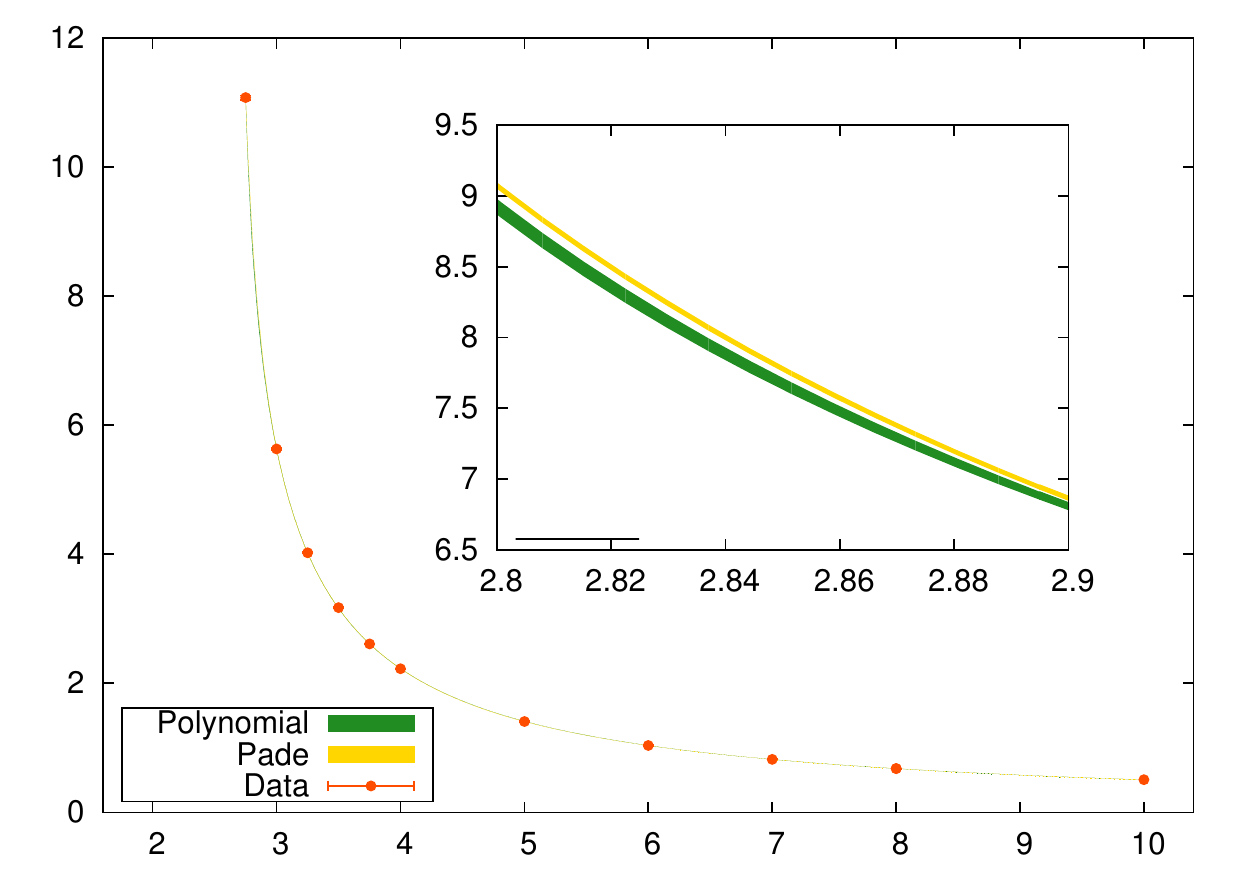}
          \caption{}
        \end{subfigure}
        \begin{subfigure}[t]{0.45\textwidth}
          \centering
          \includegraphics[width=\textwidth]{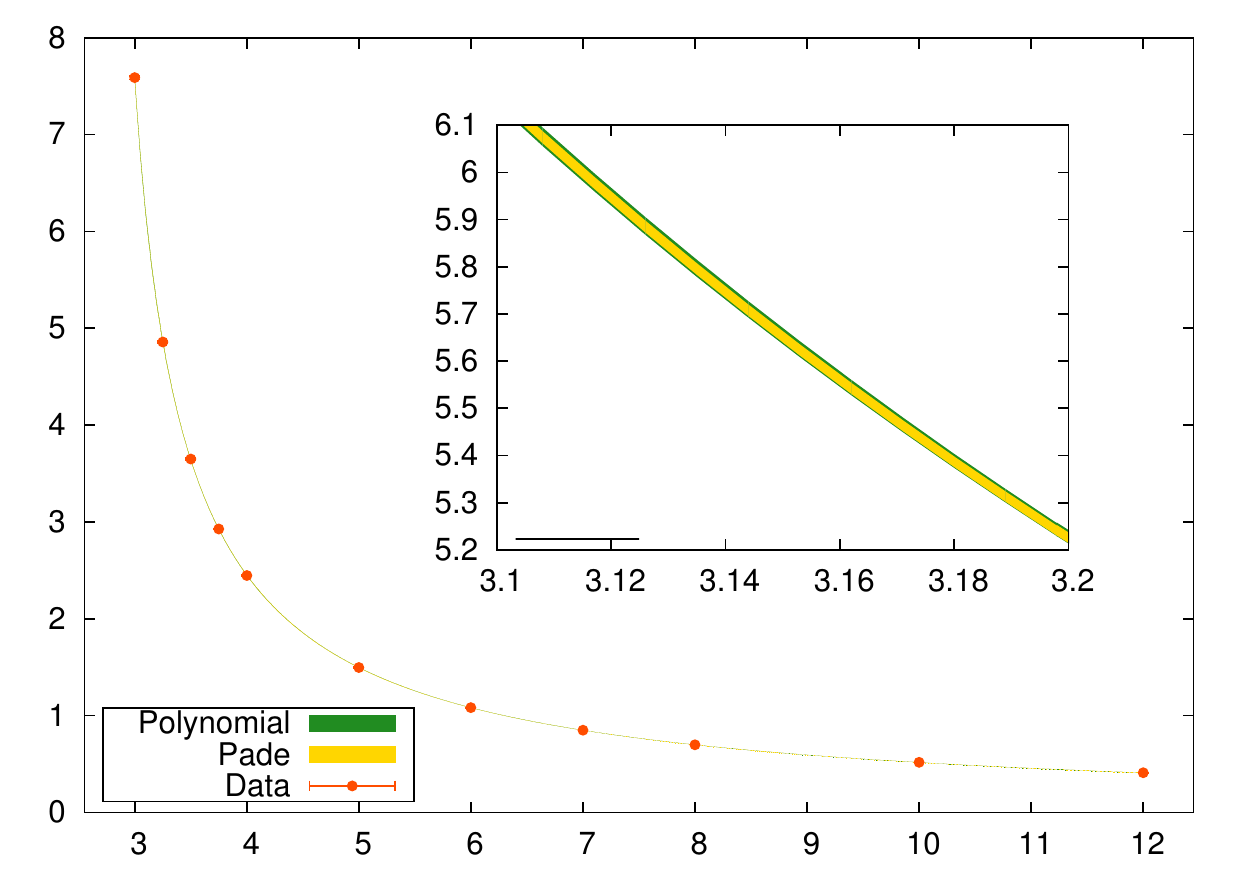}
          \caption{}
        \end{subfigure}
  \caption{Some examples of our fits to interpolate the values of the
    renormalized coupling for different values of $\beta$. 
    (a): Our worst fits corresponds to the $L/a=24$. As we
            can see there is a difference between the different
            interpolating functions between the data points. We
            stress that this systematic effects is taken into account
            in our analysis by using both functional forms to estimate
          the error of the interpolations (see the text for more details).
    (b): Fits to the data of the $L/a=36$ lattice. As we can
            see, in this case both interpolating functions agree
            within errors, although the polynomial fits tends to have
            larger errors.
}
  \label{fig:fit_l24}
\end{figure}

In the figures~\ref{fig:fit_l24} we show a couple of these fits. Our
worst fit corresponds to the $L/a=24$ lattice and 
the Pade fit gives a $\chi^2/{\rm ndof}=1.69$, while the Taylor fit
results in a fit quality of $\chi^2/{\rm ndof}=1.9$. We see how
in this case the two different functional forms interpolate
differently between the data, giving us confidence that if one
estimates the error of the interpolation using both functional forms,
one will be on the safe side\footnote{We point that probably a more
sophisticated analysis technique (or simply, simulating an additional
lattice to avoid having large gaps in the data), might result in a
more precise result.}. 

We use resampling methods to propagate errors by using 4000 bootstrap
samples. All fitting parameters derived from our original data are
computed for each bootstrap sample. Interpolation points are computed
for each bootstrap sample and each functional form. The final error of
the interpolated point is computed using \emph{both} functional forms
and \emph{all} bootstrap samples,
and therefore takes into account not only the statistical uncertainty,
but also the systematic effect due to the dependence of the
interpolating functional form. 

\subsubsection{Step scaling function}

\begin{figure}
  \centering
        \begin{subfigure}{\textwidth}
          \centering
          \includegraphics[width=\textwidth]{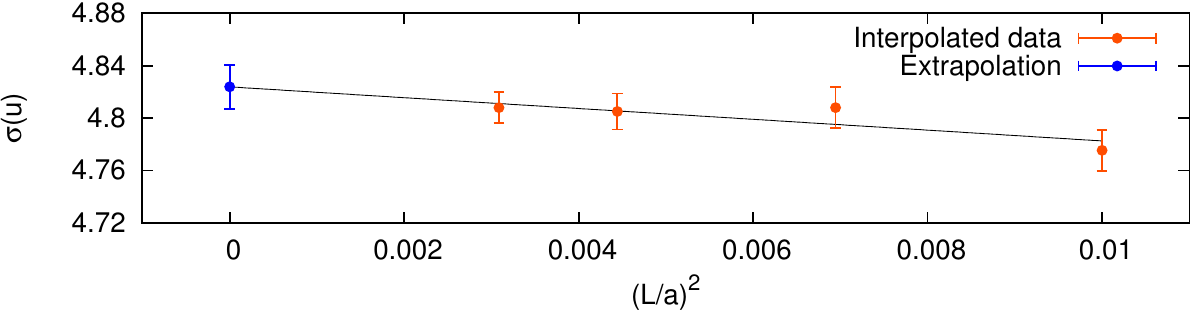}
        \end{subfigure}
        \begin{subfigure}[b]{\textwidth}
          \centering
          \includegraphics[width=\textwidth]{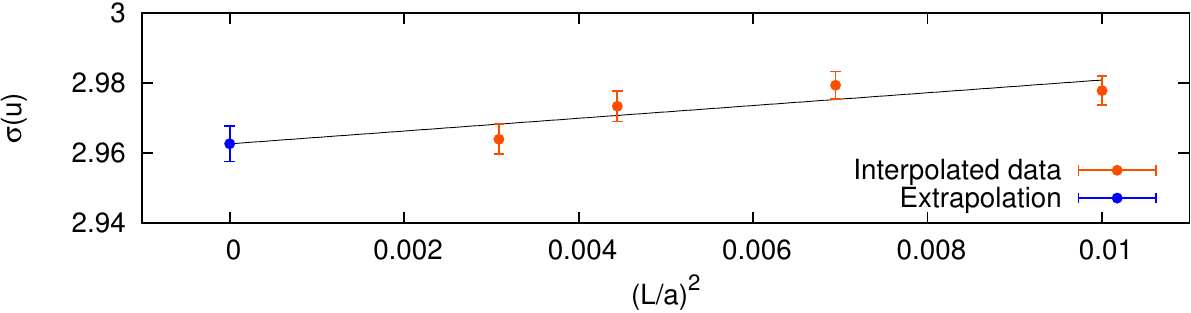}
        \end{subfigure}
        \begin{subfigure}[b]{\textwidth}
          \centering
          \includegraphics[width=\textwidth]{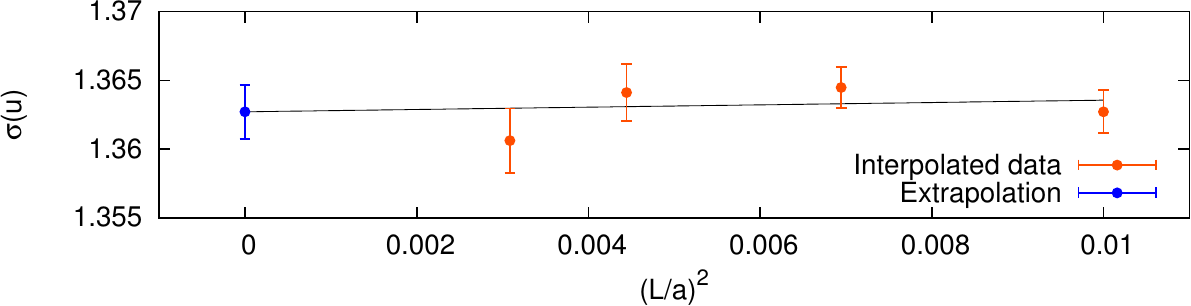}
        \end{subfigure}
  \caption{Examples of the continuum extrapolation of the step scaling
  function. The three figures corresponds (from top to bottom) to the
  values $u=7.5, 3.75, 1.5$. We recall that we use a scale factor
  $s=1/2$, and the scheme is defined by the parameter $c=0.3$.} 
  \label{fig:ss}
\end{figure}

We will first show the continuum extrapolations of the step scaling
function $\Sigma(u,a/L)$ at some representative values of
$u=7.5, 3.75, 1.5$. Figure~\ref{fig:ss} shows that these
extrapolations are mild. We have used the value $c=0.3$ that gives a
precision in the data for the renormalized coupling between $0.15\%$
and $0.25\%$.

One of the advantages of the use of the
twisted boundary conditions is the absence of $\mathcal O(a)$ cutoff
effects, that are present for example in the Schr\"odinger functional
due to boundary effects. Here the invariance under
translations guarantee that the continuum limit can be safely taken by
a linear extrapolation in $(a/L)^2$.

\subsubsection{Running coupling}

As a final application, we will compute the running coupling.
We will fix the scheme by setting 
$c=0.3$. We start our recursion in a volume $L_{\rm max}$ defined by the
condition
\begin{equation}
  g^2_{\rm TGF}(L_{\rm max})\Big|_{c=0.3} = 7.5\,.
\end{equation}
The lattice step scaling function and its continuum limit is computed
as described in the previous sections. As figure~\ref{fig:ss}
shows, the extrapolations towards the 
continuum are rather flat. The continuum limit values are used to
further compute the values of the step scaling function at larger
renormalization scales (smaller volumes), up to $L_{\rm min} = L_{\rm
  max}/2^{26}$, 
where $g^2_{\rm TGF}(L_{\rm min})|_{c=0.3}=0.5324(84)$.

Since the same functional form (fitting parameters) are used
recursively to compute the values of the coupling at different scales,
one has to propagate errors taking into account the correlations
correctly. This is done in the spirit of the resampling methods in the
most naive way: one uses as input for the coupling at a scale $L$ all
the bootstrap samples of the coupling from the scale $2L$. We recall
here that these bootstrap samples carry the information not only of
the statistical uncertainties, but also of the dependence of our
results on the functional form chosen to fit the data.
Our results have carefully taken into account the two sources of
systematic uncertainty: the continuum extrapolation and the choice of
fitting function for our lattice data.

\begin{figure}
  \centering
  \includegraphics[width=\textwidth]{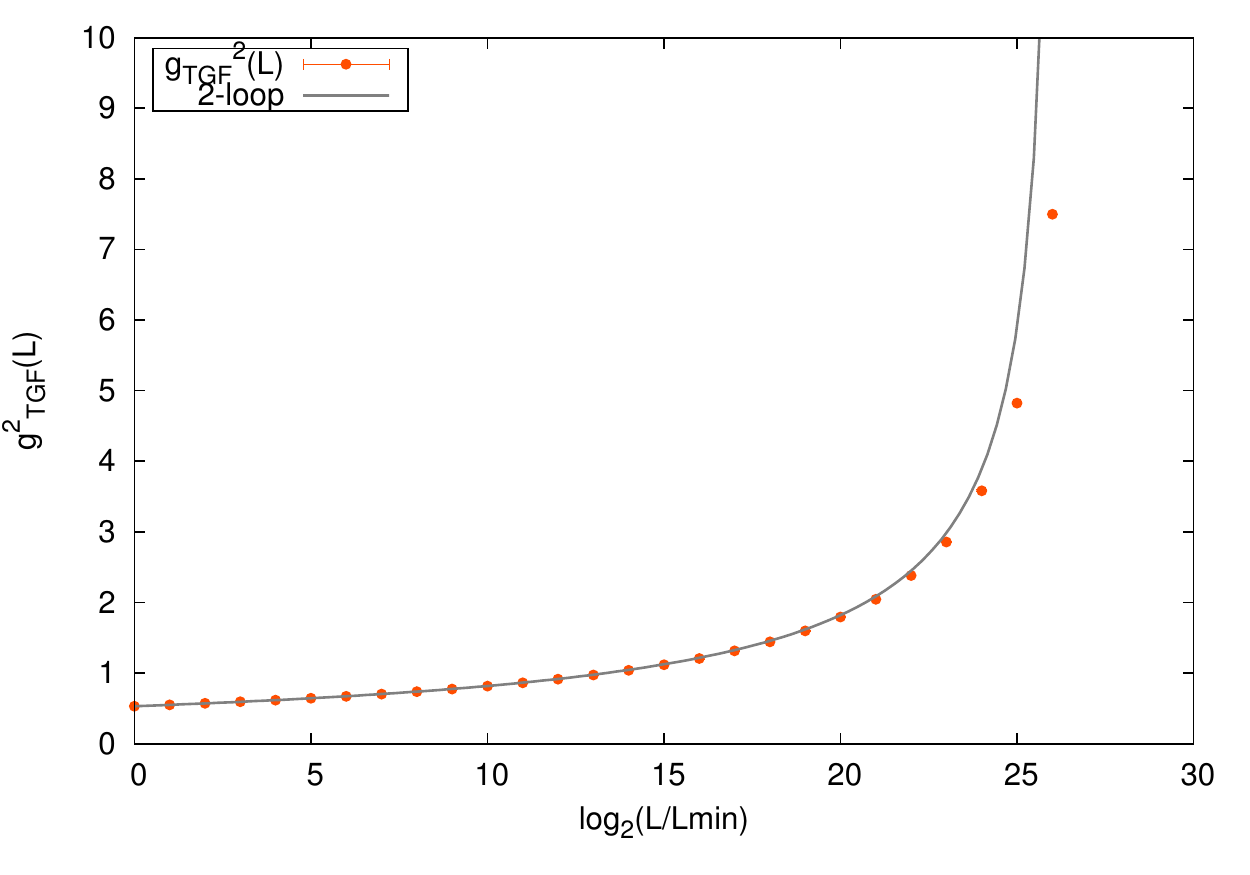}
  \caption{$g^2_{\rm TGF}(L)$ as a function of the renormalization
    scale $\log(L/L_{\rm min})$, and a comparison with the two loop
    perturbative prediction. Errors are plotted, but compatible with
    the size of the points.}
  \label{fig:gvsL}
\end{figure}

Figure~\ref{fig:gvsL} shows the running of the coupling from the low
energies to the high energies, over a factor $2^{26}$ change in scale,
while table~\ref{tab:gvsL} contains the numerical values of the
coupling at different renormalization scales. The fact that the
absolute error in the renormalized coupling tends to be constant a
large energies (small volumes), is a consequence of the error
propagation, that dominates for large energies the error budget. 

As a further consistency test, we have repeated the full running of
the coupling using as scale factor to define the step scaling function
$s=2$ (i.e. we run from high to low energies), obtaining
consistent results. 

\begin{table}
  \centering
  \begin{tabular}{l|llllll}
    \toprule
    $L=L_{\rm max}/2^k$ & $k=0$ & $k=1$ & $k=2$ & $k=3$ & $k=4$ & $k=5$ \\
    $g^2_{\rm TGF}(L)$ & 7.5 & 4.824(17) & 3.581(15) & 2.858(12) &
    2.383(10) & 2.0464(95) \\
    \midrule
    $L=L_{\rm max}/2^k$ & $k=6$ & $k=7$ & $k=8$ & $k=9$ & $k=10$ & $k=11$ \\
    $g^2_{\rm TGF}(L)$ & 1.7949(94) & 1.5995(94) & 1.4432(93) &
    1.3153(92) & 1.2085(90) & 1.1181(89) \\
    \midrule
    $L=L_{\rm max}/2^k$ & $k=12$ & $k=13$ & $k=14$ & $k=15$ & $k=16$ & $k=17$ \\
    $g^2_{\rm TGF}(L)$ & 1.0405(87) & 0.9732(86) & 0.9143(84) &
    0.8621(83) & 0.8158(83) & 0.7742(82) \\
    \midrule
    $L=L_{\rm max}/2^k$ & $k=18$ & $k=19$ & $k=20$ & $k=21$ & $k=22$ & $k=23$ \\
    $g^2_{\rm TGF}(L)$ & 0.7368(82) & 0.7028(82) & 0.6720(82) &
    0.6437(82) & 0.6178(82) & 0.5939(83) \\
    \midrule
    $L=L_{\rm max}/2^k$ & $k=24$ & $k=25$ & $k=26$ &  &  &  \\
    $g^2_{\rm TGF}(L)$ & 0.5718(83) & 0.5514(84) & 0.5324(84) &&& \\
    \bottomrule
  \end{tabular}
  \caption{Values of the renormalized twisted gradient flow coupling
    as a funtion of the renormalization scale $\mu = 1/cL$ for
    $c=0.3$. The final error at large scales
    (small volumes) is dominated by the error propagation.}
  \label{tab:gvsL}
\end{table}

The $\Lambda$ parameter can be extracted, in units of $L_{\rm max}$
via
\begin{equation}
  \Lambda = \mu(\beta_0g^2(\mu))^{-\beta_1/2\beta_0^2} e^{-1/2\beta_0g^2(\mu)}
  e^{-\int_0^{g^2(\mu)}\left\{\frac{1}{\beta(x)}+\frac{1}{\beta_0x^3}-\frac{\beta_1}{\beta_0^2x}\right\}}\,,
\end{equation}
using that $\mu = 1/cL$. The previous formula is exact, but the last
exponential is essentially unknown analytically. Nevertheless if one
uses a value of $g^2_{\rm TGF}(L)$ where the difference between the two loop
and the non-perturbative results are negligible, the effect of the
last exponential is also negligible. Of course this is
more certain the smaller the coupling, but since the relative error of
the coupling grows as the coupling decreases, this would translate in
a larger error for the $\Lambda$ parameter. Below we quote a couple of
values as example. 
\begin{eqnarray*}
  \Lambda L_{\rm max} = 1.509(44)\quad (@ g_{TGF}^2(L) = 1.7949(94))\,,\\
  \Lambda L_{\rm max} = 1.57(13)\quad  (@ g_{TGF}^2(L) = 1.0405(87))\,.
\end{eqnarray*}

We want to end this section with a small comment on the use of
different values of $c$. The main point has already been
raised in~\cite{Fritzsch:2013je}: the larger the value of $c$, the larger
the (relative) statistical error of the coupling, but the scaling
towards the continuum seems better. This general behavior is
consistent with the leading order in perturbation theory as we have
seen. We will simply say that the relative error in the raw data
increases with $c$, and roughly one can say that for $c=0.4$ the
relative error is two times larger than for $c=0.3$, while for
$c=0.5$ the error is three times larger. This statement seem to hold
true independently of the volume (i.e. of the value of $g^2_{\rm TGF}$). 


\section{Conclusions}

In this work we have studied the YM gradient flow for fields obeying 
twisted boundary conditions on a 4D torus. The energy density of the
flow field at 
positive flow time is used to define a non-perturbative coupling at a
scale given by the flow time. Moreover one can make the flow time
scale with the finite size of the box. In this way one obtains a
coupling that depends only on a scale given by the size of the torus,
and the usual techniques of step scaling can be applied. 

The perturbative behavior of the energy density of the flow field is
computed to leading order both in the lattice and in the
continuum. This allows us to estimate the size of cutoff effects of
the coupling to leading order in perturbation theory. Results are
consistent with other definitions of the running coupling using the
gradient flow. Cutoff effects are mild. 

In order to state the validity of this coupling definition beyond
perturbation theory we have performed a numerical study in pure gauge
$SU(2)$. We have computed the running of the coupling from the very
perturbative regime $g^2\sim 0.5$ to the non perturbative one $g^2\sim
7.5$, over a change by a factor $2^{26}$ in scale. We have used
lattices of sizes $L/a=10-36$. The statistical precision that one can
achieve with this coupling definition is very high: 2048 independent
measurements are enough to achieve a sub-percent precision
($0.15-0.25\%$). This precision is roughly independent of the physical
volume, the value of the coupling or the lattice size. We have also
shown that  
the continuum extrapolations of the step scaling function are rather
flat. Moreover we have shown that this techniques can be used to
reliably extract the $\Lambda$ parameter. 

The same coupling definition can be used if fermions are coupled to
the gauge field. But the consistency of the twisted boundary
conditions imposes a constraint on the number of fermions in the
fundamental representation that we can use. In particular, what 
arguably is the most interesting application of these techiniques,
namely the coupling constant of QCD (gauge group $SU(3)$ with $N_f=4$ fermions
in the fundamental representation) can not be attacked with this
method. Nevertheless there are many applications of this work. On one
hand $N_f=3$ flavour of quarks in the fundamental representation is
already interesting for the strong interations. But it
is probably in the context of conformal theories and physics beyond the
standard model, with models like $SU(3)$ with $N_f=12$ fermions in
the fundamental representation, or with adjoint fermions, where
the nice properties of this coupling definition (automatic $\mathcal
O(a)$ improvement, analyticity and high statistical precision) can
have a higher impact. In particular the setup with twisted boundary
conditions have already been used for step scaling studies in the TPL
scheme~\cite{Lin:2012iw,Itou:2012qn}. In this particular case the
use of the gradient flow observable will lead to more
precise results, as we have already seen in some preliminary results in the
last lattice conference~\cite{linconf}. One can also consider the
application of the same ideas to other choices of twisted boundary
conditions. This is specially interesting in the context of reduced
models, as we have also seen recently~\cite{liamconf}.


\section*{Acknowledgments}

This work has a large debt with
M. Garc\'ia Perez and A. Gonz\'alez-Arroyo for sharing some of their results
and notes before publication and for the many illuminating
discussions. The help and advice of R. Sommer and U. Wolff was
invaluable in many of the steps of this work. 

I also want to thank my colleagues at DESY/HU, specially
P. Korcyl, P. Fritzsch, S. Sint and R. Sommer for the very many interesting
discussions and their careful reading of the manuscript. D. Lin was very
kind reading and helping to improve a manuscript of this work. 

\appendix

\section{Raw values of $g_{TGF}^2$ }
\label{ap:values}

All the values quoted in this appendix are computed using
Eq.~\ref{eq:latcou}. Simulations are performed as described in
section~\ref{sc:run}.

\begin{table}
  \centering
  \begin{tabular}{l|lllll}
    \toprule
    $\beta$ & 10 & 12 & 15 & 18 & $L/a$\\
    \midrule
    12.0 & $0.39194(60)$ & $0.39465(61)$ & $0.39813(62)$ &$0.39974(59)$ & \\
    10.0 & $0.48769(76)$ & $0.49211(74)$ & $0.49688(76)$ &$0.49945(76)$ & \\
    8.0 &  $0.6455(10)$  & $0.6517(10)$  & $0.6605(10)$ & $0.6693(10)$ & \\
    7.0 &  $0.7701(12)$  & $0.7824(12)$  & $0.7945(13)$ & $0.8055(13)$ & \\
    6.0 &  $0.9574(15)$  & $0.9738(16)$  & $0.9934(16)$ & $1.0087(16)$ & \\
    5.0 &  $1.2644(20)$  & $1.2933(22)$  & $1.3269(22)$ & $1.3573(22)$ & \\
    4.0 &  $1.8747(31)$  & $1.9390(33)$  & $2.0194(36)$ & $2.0883(36)$& \\
    3.75 & $2.1336(38)$  & $2.2138(39)$  & $2.3259(41)$ & $2.4263(43)$& \\
    3.5 &  $2.4910(44)$  & $2.5971(47)$  & $2.7584(52)$ & $2.8999(53)$& \\
    3.25 & $2.9803(55)$  & $3.1476(59)$  & $3.4002(66)$ & $3.6192(70)$& \\
    3.0 &  $3.7453(73)$  & $4.0287(79)$  & $4.4214(87)$ & $4.8287(99)$& \\
    2.75 & $5.119(11)$   & $5.684(13)$   & $6.622(16)$  & $7.745(21) $& \\
    \bottomrule
  \end{tabular}
  \centering
  \begin{tabular}{l|lllll}
    \toprule
    $\beta$ & 20 & 24 & 30 & 36 & $L/a$\\
    \midrule
    12.0 & $0.40162(60)$ & $-$         &$0.40772(61)$ &$0.41078(63)$ & \\
    10.0 & $0.50340(76)$ &$0.50786(77)$&$0.51280(82)$ &$0.51809(85)$ & \\
    8.0 &  $0.6741(11)$ &  $0.6794(11)$&$0.6900(11)$  &$0.6987(11)$ & \\
    7.0 &  $0.8125(13)$ &  $0.8240(13)$&$0.8369(13)$  &$0.8497(13)$ & \\
    6.0 &  $1.0222(17)$ &  $1.0379(17)$&$1.0609(17)$  &$1.0819(18)$ & \\
    5.0 &  $1.3767(23)$ &  $1.4091(23)$&$1.4562(25)$  &$1.4968(25)$ & \\
    4.0 &  $2.1439(39)$ &  $2.2260(41)$&$2.3453(42)$  &$2.4465(43)$ & \\
    3.75 & $2.5047(47)$ &  $2.6107(48)$&$2.7636(52)$  &$2.9277(54)$ & \\
    3.5 &  $3.0037(57)$ &  $3.1720(60)$&$3.4170(66)$  &$3.6494(70)$ & \\
    3.25 & $3.7581(75)$ &  $4.0224(75)$&$4.4397(86)$  &$4.8568(98)$ & \\
    3.0 &  $5.088(11)$ &   $5.630(12)$ &$6.573(16)$   &$7.587(20)$ & \\
    2.75 & $8.699(25)$ &   $11.072(34)$ &$15.817(44)$ &$-$ & \\
    \bottomrule
  \end{tabular}
  \caption{Raw values of $g^2_{\rm TGF}(\beta;L/a)$.}
  \label{tab:raw_double}
\end{table}


\bibliography{/home/alberto/docs/bib/math,/home/alberto/docs/bib/campos,/home/alberto/docs/bib/fisica,/home/alberto/docs/bib/computing}

\providecommand{\href}[2]{#2}\begingroup\raggedright\begin{thebibliography}{10}

\bibitem{PhysRevLett.30.1343}
D.~J. Gross and F.~Wilczek, {\it Ultraviolet behavior of non-abelian gauge
  theories},  {\em Phys. Rev. Lett.} {\bf 30} (Jun, 1973) 1343--1346.

\bibitem{PhysRevLett.30.1346}
H.~D. Politzer, {\it Reliable perturbative results for strong interactions?},
  {\em Phys. Rev. Lett.} {\bf 30} (Jun, 1973) 1346--1349.

\bibitem{Luscher:1991wu}
M.~L{\"u}scher, P.~Weisz, and U.~Wolff, {\it {A Numerical method to compute the
  running coupling in asymptotically free theories}},  {\em Nucl.Phys.} {\bf
  B359} (1991) 221--243.

\bibitem{Luscher:1992an}
M.~L{\"u}scher, R.~Narayanan, P.~Weisz, and U.~Wolff, {\it {The Schr{\"o}dinger
  Functional: a renormalizable probe for non-abelian gauge theories}},  {\em
  Nucl.Phys.} {\bf B384} (1992) 168--228,
  [\href{http://xxx.lanl.gov/abs/hep-lat/9207009}{{\tt hep-lat/9207009}}].

\bibitem{deDivitiis:1994yp}
G.~de~Divitiis, R.~Frezzotti, M.~Guagnelli, and R.~Petronzio, {\it
  {Non-perturbative determination of the running coupling constant in quenched
  SU(2)}},  {\em Nucl.Phys.} {\bf B433} (1995) 390--402,
  [\href{http://xxx.lanl.gov/abs/hep-lat/9407028}{{\tt hep-lat/9407028}}].

\bibitem{Luscher:1992zx}
M.~Luscher, R.~Sommer, U.~Wolff, and P.~Weisz, {\it {Computation of the running
  coupling in the SU(2) Yang-Mills theory}},  {\em Nucl.Phys.} {\bf B389}
  (1993) 247--264, [\href{http://xxx.lanl.gov/abs/hep-lat/9207010}{{\tt
  hep-lat/9207010}}].

\bibitem{Luscher:1993gh}
M.~L{\"u}scher, R.~Sommer, P.~Weisz, and U.~Wolff, {\it {A precise
  determination of the running coupling in the SU(3) Yang-Mills theory}},  {\em
  Nucl.Phys.} {\bf B413} (1994) 481--502,
  [\href{http://xxx.lanl.gov/abs/hep-lat/9309005}{{\tt hep-lat/9309005}}].

\bibitem{DellaMorte:2004bc}
{\bf ALPHA} Collaboration, M.~Della~Morte et~al., {\it {Computation of the
  strong coupling in QCD with two dynamical flavors}},  {\em Nucl.Phys.} {\bf
  B713} (2005) 378--406, [\href{http://xxx.lanl.gov/abs/hep-lat/0411025}{{\tt
  hep-lat/0411025}}].

\bibitem{Tekin:2010mm}
{\bf ALPHA} Collaboration, F.~Tekin, R.~Sommer, and U.~Wolff, {\it {The running
  coupling of QCD with four flavors}},  {\em Nucl.Phys.} {\bf B840} (2010)
  114--128, [\href{http://xxx.lanl.gov/abs/1006.0672}{{\tt arXiv:1006.0672}}].

\bibitem{Sint:1993un}
S.~Sint, {\it {On the Schr{\"o}dinger functional in QCD}},  {\em Nucl.Phys.}
  {\bf B421} (1994) 135--158,
  [\href{http://xxx.lanl.gov/abs/hep-lat/9312079}{{\tt hep-lat/9312079}}].

\bibitem{Sint:1995rb}
S.~Sint, {\it {One loop renormalization of the QCD Schr\"odinger functional}},
  {\em Nucl.Phys.} {\bf B451} (1995) 416--444,
  [\href{http://xxx.lanl.gov/abs/hep-lat/9504005}{{\tt hep-lat/9504005}}].

\bibitem{Sint:2010eh}
S.~Sint, {\it {The Chirally rotated Schr\"odinger functional with Wilson
  fermions and automatic O(a) improvement}},  {\em Nucl.Phys.} {\bf B847}
  (2011) 491--531, [\href{http://xxx.lanl.gov/abs/1008.4857}{{\tt
  arXiv:1008.4857}}].

\bibitem{Frezzotti:2003ni}
R.~Frezzotti and G.~Rossi, {\it {Chirally improving Wilson fermions. 1. O(a)
  improvement}},  {\em JHEP} {\bf 0408} (2004) 007,
  [\href{http://xxx.lanl.gov/abs/hep-lat/0306014}{{\tt hep-lat/0306014}}].

\bibitem{Luscher:2010iy}
M.~L{\"u}scher, {\it {Properties and uses of the Wilson flow in lattice QCD}},
  {\em JHEP} {\bf 1008} (2010) 071,
  [\href{http://xxx.lanl.gov/abs/1006.4518}{{\tt arXiv:1006.4518}}].

\bibitem{Luscher:2011bx}
M.~L{\"u}scher and P.~Weisz, {\it {Perturbative analysis of the gradient flow
  in non-abelian gauge theories}},  {\em JHEP} {\bf 1102} (2011) 051,
  [\href{http://xxx.lanl.gov/abs/1101.0963}{{\tt arXiv:1101.0963}}].

\bibitem{Fodor:2012td}
Z.~Fodor, K.~Holland, J.~Kuti, D.~Nogradi, and C.~H. Wong, {\it {The Yang-Mills
  gradient flow in finite volume}},  {\em JHEP} {\bf 1211} (2012) 007,
  [\href{http://xxx.lanl.gov/abs/1208.1051}{{\tt arXiv:1208.1051}}].

\bibitem{GonzalezArroyo:1981vw}
A.~Gonzalez-Arroyo, J.~Jurkiewicz, and C.~Korthals-Altes, {\it {Ground state
  metamorphosis for Yang-Mills fields on a finite periodic lattice}}, .

\bibitem{vanBaal:1988qm}
P.~van Baal, {\it {Gauge theory in a finite volume}},  {\em Acta Phys.Polon.}
  {\bf B20} (1989) 295--312.

\bibitem{Fritzsch:2013je}
P.~Fritzsch and A.~Ramos, {\it {The gradient flow coupling in the Schrödinger
  Functional}},  {\em JHEP} {\bf 1310} (2013) 008,
  [\href{http://xxx.lanl.gov/abs/1301.4388}{{\tt arXiv:1301.4388}}].

\bibitem{Schaefer:2010hu}
{\bf ALPHA} Collaboration, S.~Schaefer, R.~Sommer, and F.~Virotta, {\it
  {Critical slowing down and error analysis in lattice QCD simulations}},  {\em
  Nucl. Phys.} {\bf B845} (2011) 93--119,
  [\href{http://xxx.lanl.gov/abs/1009.5228}{{\tt arXiv:1009.5228}}].

\bibitem{Luscher:2011kk}
M.~L{\"u}scher and S.~Schaefer, {\it {Lattice QCD without topology barriers}},
  {\em JHEP} {\bf 1107} (2011) 036,
  [\href{http://xxx.lanl.gov/abs/1105.4749}{{\tt arXiv:1105.4749}}].

\bibitem{Fritzsch:2013yxa}
P.~Fritzsch, A.~Ramos, and F.~Stollenwerk, {\it {Critical slowing down and the
  gradient flow coupling in the Schr\"odinger functional}},  {\em PoS} {\bf
  Lattice2013} (2013) 461, [\href{http://xxx.lanl.gov/abs/1311.7304}{{\tt
  arXiv:1311.7304}}].

\bibitem{Luscher:2014kea}
M.~Lüscher, {\it {Step scaling and the Yang-Mills gradient flow}},  {\em JHEP}
  {\bf 1406} (2014) 105, [\href{http://xxx.lanl.gov/abs/1404.5930}{{\tt
  arXiv:1404.5930}}].

\bibitem{Luscher:2012av}
M.~L{\"u}scher and S.~Schaefer, {\it {Lattice QCD with open boundary conditions
  and twisted-mass reweighting}},  {\em Comput.Phys.Commun.} {\bf 184} (2013)
  519--528, [\href{http://xxx.lanl.gov/abs/1206.2809}{{\tt arXiv:1206.2809}}].

\bibitem{ga:torus}
A.~Gonz{\'a}lez-Arroyo, {\it {Yang-Mills fields on the 4-dimensional Torus.
  Part I: Classical Theory}},  {\em {World Scientific. Proceedings of the
  Pe\~niscola 1997 advanced school on non-perturbative quantum field physics}}
  (1998) {Singapore}, [\href{http://xxx.lanl.gov/abs/hep-th/9807108}{{\tt
  hep-th/9807108}}].

\bibitem{Perez:2013dra}
M.~Garc\'ia~P\'erez, A.~Gonz\'alez-Arroyo, and M.~Okawa, {\it {Spatial volume
  dependence for 2+1 dimensional SU(N) Yang-Mills theory}},  {\em JHEP} {\bf
  1309} (2013) 003, [\href{http://xxx.lanl.gov/abs/1307.5254}{{\tt
  arXiv:1307.5254}}].

\bibitem{tHooft:1979uj}
G.~'t~Hooft, {\it A property of electric and magnetic flux in nonabelian gauge
  theories},  {\em Nucl. Phys.} {\bf B153} (1979) 141.

\bibitem{Luscher:1985wf}
M.~Luscher and P.~Weisz, {\it {Efficient Numerical Techniques for Perturbative
  Lattice Gauge Theory Computations}},  {\em Nucl.Phys.} {\bf B266} (1986) 309.

\bibitem{Parisi:1984cy}
G.~Parisi, {\it {Prolegomena to any future computer evaluation of the QCD mass
  spectrum}},  {\em {Cargese Summer Inst. 1983:0531}} (1984).

\bibitem{Narayanan:2006rf}
R.~Narayanan and H.~Neuberger, {\it {Infinite N phase transitions in continuum
  Wilson loop operators}},  {\em JHEP} {\bf 0603} (2006) 064,
  [\href{http://xxx.lanl.gov/abs/hep-th/0601210}{{\tt hep-th/0601210}}].

\bibitem{Luscher:2009eq}
M.~L{\"u}scher, {\it {Trivializing maps, the Wilson flow and the HMC
  algorithm}},  {\em Commun.Math.Phys.} {\bf 293} (2010) 899--919,
  [\href{http://xxx.lanl.gov/abs/0907.5491}{{\tt arXiv:0907.5491}}].

\bibitem{Luscher:2013cpa}
M.~Luscher, {\it {Chiral symmetry and the Yang--Mills gradient flow}},  {\em
  JHEP} {\bf 1304} (2013) 123, [\href{http://xxx.lanl.gov/abs/1302.5246}{{\tt
  arXiv:1302.5246}}].

\bibitem{Creutz:1980zw}
M.~Creutz, {\it {Monte Carlo Study of Quantized SU(2) Gauge Theory}},  {\em
  Phys.Rev.} {\bf D21} (1980) 2308--2315.

\bibitem{Fabricius:1984wp}
K.~Fabricius and O.~Haan, {\it {Heat Bath Method for the Twisted {Eguchi-Kawai}
  Model}},  {\em Phys.Lett.} {\bf B143} (1984) 459.

\bibitem{Kennedy:1985nu}
A.~Kennedy and B.~Pendleton, {\it {Improved Heat Bath Method for Monte Carlo
  Calculations in Lattice Gauge Theories}},  {\em Phys.Lett.} {\bf B156} (1985)
  393--399.

\bibitem{Creutz:1987xi}
M.~Creutz, {\it {Overrelaxation and Monte Carlo Simulation}},  {\em Phys.Rev.}
  {\bf D36} (1987) 515.

\bibitem{Wolff:1992nq}
U.~Wolff, {\it {Dynamics of hybrid overrelaxation in the Gaussian model}},
  {\em Phys.Lett.} {\bf B288} (1992) 166--170.

\bibitem{Lin:2012iw}
C.-J.~D. Lin, K.~Ogawa, H.~Ohki, and E.~Shintani, {\it {Lattice study of
  infrared behaviour in SU(3) gauge theory with twelve massless flavours}},
  {\em JHEP} {\bf 1208} (2012) 096,
  [\href{http://xxx.lanl.gov/abs/1205.6076}{{\tt arXiv:1205.6076}}].

\bibitem{Itou:2012qn}
E.~Itou, {\it {Properties of the twisted Polyakov loop coupling and the
  infrared fixed point in the SU(3) gauge theories}},  {\em PTEP} {\bf 2013}
  (2013), no.~8 083B01, [\href{http://xxx.lanl.gov/abs/1212.1353}{{\tt
  arXiv:1212.1353}}].

\bibitem{linconf}
D.~Lin, {\it {SU(3) gauge theory with 12 flavours in a twisted box}.},  {\em
  Talk at The 32nd International Symposium on Lattice Field Theory} (2014).

\bibitem{liamconf}
L.~Keegan, {\it {TEK twisted gradient flow running coupling}.},  {\em Talk at
  The 32nd International Symposium on Lattice Field Theory} (2014).

\end{thebibliography}\endgroup

\end{document}